\documentclass[fleqn,usenatbib]{mnras}

\usepackage{amsmath}
\usepackage{txfonts}                
\usepackage{graphicx}               
\usepackage[T1]{fontenc}
\hypersetup{linkcolor=red}          

\usepackage{enumerate}
\usepackage{hyperref}

\newcommand{\aref}[1]{\hyperref[#1]{Appendix~\ref{#1}}}

\title[kinematic-morphology vs star-formation relation]{SDSS-IV MaNGA: The kinematic-morphology of galaxies on the mass vs star-formation relation in different environments}

\author[B. Wang et al.]{
Bitao Wang,$^{1,2,3}$\thanks{E-mail: bt-wang@pku.edu.cn}
Michele Cappellari,$^{3}$
Yingjie Peng,$^{2}$\thanks{E-mail: yjpeng@pku.edu.cn}
Mark Graham$^{3}$
\\
$^{1}$Department of Astronomy, School of Physics, Peking University, Beijing 100871, China\\
$^{2}$Kavli Institute for Astronomy and Astrophysics, Peking University, Beijing 100871, China\\
$^{3}$Sub-department of Astrophysics, Department of Physics, University of Oxford, Denys Wilkinson Building, Keble Road, Oxford OX1 3RH, UK
}

\date{Accepted for publication on MNRAS on 2020 May 07}

\pubyear{2020}

\begin{document}
\label{firstpage}
\pagerange{\pageref{firstpage}--\pageref{lastpage}}
\maketitle

\begin{abstract}

We study the link between the kinematic-morphology of galaxies, as inferred from integral-field stellar kinematics, and their relation between mass and star formation rate. Our sample consists of $\sim 3200$ galaxies with integral-field spectroscopic data from the MaNGA survey (Mapping Nearby Galaxies at Apache Point Observatory) with available determinations of their effective stellar angular momentum within the half-light radius $\lambda_{R_e}$. We find that for star-forming galaxies, namely along the star formation main sequence (SFMS), the $\lambda_{R_e}$ values remain large and almost unchanged over about two orders of magnitude in stellar mass, with the exception of the lowest masses $\mathcal{M}_{\star}\lesssim2\times10^{9} \mathcal{M}_{\odot}$, where $\lambda_{R_e}$ slightly decreases. The SFMS is dominated by spiral galaxies with small bulges. Below the SFMS, but above the characteristic stellar mass $\mathcal{M}_{\rm crit}\approx2\times10^{11} \mathcal{M}_{\odot}$, there is a sharp decrease in $\lambda_{R_e}$ with decreasing star formation rate (SFR): massive galaxies well below the SFMS are mainly slow-rotator early-type galaxies, namely genuinely spheroidal galaxies without disks. Below the SFMS and below $\mathcal{M}_{\rm crit}$ the decrease of $\lambda_{R_e}$ with decreasing SFR becomes modest or nearly absent: low-mass galaxies well below the SFMS, are fast-rotator early-type galaxies, and contain fast-rotating stellar disks like their star-forming counterparts. We also find a small but clear environmental dependence for the massive galaxies: in the mass range $10^{10.9}-10^{11.5} \mathcal{M}_{\odot}$, galaxies in rich groups or denser regions or classified as central galaxies have lower values of $\lambda_{R_e}$. While no environmental dependence is found for galaxies of lower mass. We discuss how the above results can be understood as due to the different star formation and mass assembly histories of galaxies with varying mass.

\end{abstract}

\begin{keywords}
galaxies:evolution--galaxies:formation--galaxies:kinematics and dynamics
\end{keywords}



\section{Introduction}

The diverse colour and morphology are probably the most two striking and straightforward features of galaxies.
Roughly at the time when morphological classification systems more refined than the well-known Hubble scheme came out \citep[e.g.,][]{1959HDP....53..275D, 1960ApJ...131..215V}, astronomers started to establish that these two properties of nearby galaxies actually correlate with each other in the sense that elliptical galaxies are typically red whereas spiral galaxies are bluer \citep[e.g.,][]{1958MeLuS.136....1H}.

In the following years, the advent of larger telescopes and the use of charged coupled devices allowed for detailed study of light profiles \citep{1963BAAA....6...41S} and galaxy structure components such as bulges, disks, bars and rings \citep[e.g.,][]{1977ApJ...217..406K, 1991rc3..book.....D, 1993MNRAS.265.1013C}.
In parallel, the accumulation of stellar spectra deepened our understanding of galaxy integrated light and their stellar populations \citep[e.g.,][]{1983ApJ...273..105B, 1985ApJS...57..711F, 1994ApJS...94..687W}.

But not until the new millennium had our knowledge about this morphology-colour relation (hereafter T-C relation), i.e. ``elliptical galaxies are red and spiral galaxies are blue'', been dramatically updated.
Large sky surveys such as The Sloan Digital Sky Survey (SDSS) revealed the sharp bimodality of low-z galaxies on colour-magnitude diagram \citep{2001AJ....122.1861S, 2004ApJ...600..681B} which evolves to the form of two sequences of star-forming and quiescent/passive galaxies\footnote{Though the passive sequence sometimes is less obvious as its correspondence red sequence on colour-magnitude diagram because SFR does not necessarily saturates to the extent as much as colour index.}
on a diagram of two more physical parameters, the star formation rate-stellar mass ($\mathrm{SFR}-\mathcal{M}_{\star}$) diagram \citep{2004MNRAS.351.1151B, 2007ApJ...660L..43N}.
Also thanks to observations at higher redshift $z$ which clearly show the growth of passive population \citep[e.g.,][]{2004ApJ...608..752B, 2007ApJ...665..265F}, astronomers started picturing galaxy evolution tracks in this diagram.
The ``star formation main sequence'' (SFMS) consists of actively star-forming galaxies.
For some reasons some of these leave this sequence (the so-called quenching of star formation) and finally join passive population \citep{2007ApJS..173..342M, 2007ApJS..173..267S, 2007ApJS..173..315S, 2012MNRAS.420.1684W}.
Between the red sequence and the blue cloud lies the "green valley" which is made by galaxies believed to be transitioning from one sequence to the other.
Therefore it has received much attention because of its potential for shedding light on the exact reasons why galaxies leave SFMS \citep[e.g.,][]{2011ApJ...736..110M, 2014MNRAS.440..889S, 2014SerAJ.189....1S}.
Being a hot topic, much effort has been made to understand galaxy evolution on this $\mathrm{SFR}-\mathcal{M}_{\star}$ diagram and the quenching of star formation.
And by looking for galaxy properties closest to star formation state, that T-C relation regained its significance and was presented in a modern look.

It has been extensively reported that the structure (in terms of light concentration, bulge to total light ratio, S{\'e}rsic index, central density etc.) of galaxies and their star formation state (indicated by the position on colour-magnitude or $\mathrm{SFR}-\mathcal{M}_{\star}$ diagram) are closely related in the same sense as that original T-C relation, both at low $z$ \citep[e.g.,][]{2003MNRAS.341...54K, 2009ApJ...699..105C, 2009MNRAS.393.1531G,  2019MNRAS.485..666B} and at higher z \citep[e.g.,][]{2011ApJ...742...96W, 2012ApJ...753..167B, 2012ApJ...760..131C, 2014ApJ...788...11L, 2017ApJ...838...19W}.
These results all showed that structure is a good indicator of star formation rate and suggested a quenching mechanism connected to morphology.
For example, a scenario has been proposed where a massive bulge can stabilise the disk against fragmentation so that cessation of star formation is achieved \citep{2009ApJ...707..250M}.
And less directly, as bulge mass and the mass of galactic super massive black hole (SMBH) are connected via $M-\sigma$ relation \citep{2013ARA&A..51..511K}, the influence from SMBH is also considered a driver of the T-C relation.
Wet compactions induced by, e.g., mergers, counter rotating streams or infall of galactic fountain raise mass concentration as well as feed the central SMBH, and it is followed by quenching in the form of stellar and active galactic nucleus (AGN) feedback resulting in a compact and passive galaxy \citep{2014MNRAS.438.1870D, 2015MNRAS.450.2327Z, 2016MNRAS.457.2790T, 2019arXiv190408431D}.
Regardless of whether bulges and AGNs directly quench the entire galaxies or not (which so far have not been unambiguously observed), it has been reported that bulges have lower specific star formation rate than disks \citep[e.g.,][]{2014ApJ...785L..36A, 2017ApJ...851...18L}.

Morphology reflects the orbital composition of a galaxy.
Galaxy spheroids have low angular momentum and higher random motions while disky galaxies are dominated by stars in ordered rotation.
By studying the kinematic properties of galaxies we get complementary and sometimes revolutionary results for light distribution based studies.
As opposed to what people originally thought, long-slit spectragraphs revealed that fainter elliptical galaxies have higher degree of rotation than the brighter ones \citep{1975ApJ...200..439B, 1977ApJ...218L..43I, 1982ApJ...256..460K, 1982ApJ...257...75K, 1983ApJ...266...41D}.
The advent of integral-field spectroscopy (IFS) with the SAURON \citep{2002MNRAS.329..513D}, suggested a possible dichotomy between the kinematics of these two classes of galaxies that they called fast and slow rotators early-type galaxies \citep{2007MNRAS.379..401E, 2007MNRAS.379..418C}.
This dichotomy appeared closely related to the suggested one based on the inner surface brightness profiles \citep{1996ApJ...464L.119K, 1997AJ....114.1771F}.
Later, the volume-limited $\mathrm{ATLAS}^{\mathrm{3D}}$ IFS survey \citep{2011MNRAS.413..813C} provided a proper census of the kinematic morphology in galaxies \citep{2011MNRAS.414.2923K, 2011MNRAS.414..888E}.
More recently, the evidence for a dichotomy was placed on firm ground \citep{2018MNRAS.477.4711G} using data from the Mapping Nearby Galaxies at Apache Point Observatory (MaNGA) survey \citep{2015ApJ...798....7B}.
See \citet{2016ARA&A..54..597C} for a review.
This kinematic census has drawn attention to the similarity between intrinsic morphology of spiral galaxies and fast rotators, and has led to a revision of the traditional Hubble morphological classification scheme which ignores the wide range of bulge size of S0s \citep{2011MNRAS.416.1680C}.

Part of the breakthrough is because classifications and quantifications based on kinematic maps are less affected by projection.
And this can also be taken as advantage to investigate T-C relation.
Current studies have shown the difference between fast and slow rotators in terms of current star formation and star formation histories \citep[e.g.,][]{2010MNRAS.402.2140S, 2018MNRAS.473.2679S}.
And echoing those studies suggesting the significance of central density in quenching galaxies, \citet{2013MNRAS.432.1862C} and \citet{2018MNRAS.476.1765L} found the important role of velocity dispersion in driving the variation of stellar populations both of early-type and late-type galaxies.
A full kinematic version of the T-C relation has been shown in \citet{2018NatAs...2..483V}.
The intrinsic shape of galaxies, as quantified in $V/\sigma - \epsilon$ (ordered over random motion versus ellipticity) diagram with certain assumptions, clearly correlates with their luminosity-weighted age in the results.

In this work, we attempt to link together the information in the $\mathrm{SFR} - \mathcal{M}_{\star}$ and $(\lambda,\epsilon)$ diagrams, two extensively exploited tools for understanding the star formation and kinematic aspects of galaxies in the literature.
The latter, firstly introduced in \citet{2007MNRAS.379..401E} as an improved version of the $(V/\sigma,\epsilon)$ (with more introduction in \autoref{sec:measure}), takes into account the spatial structure of kinematic maps and has been shown able to differentiate between kinematic classes \citep{2016ARA&A..54..597C, 2018MNRAS.477.4711G}.
In conjunction, we are able to revisit T-C relation from a new and informative perspective.
And by studying $\sim 3200$ galaxies with IFS data in data release 15 of Mapping Nearby Galaxies at Apache Point Observatory (MaNGA) survey \citep{2015ApJ...798....7B}, the largest IFS sample up-to-date, we bring considerable statistical significance in our work.
Such, $\mathrm{SFR} - \mathcal{M}_{\star}$ diagram can be well covered down to a stellar mass $\sim 10^{9.5}\,\mathcal{M}_{\odot}$.
Special attention will be paid to the dependence on stellar mass and environment, which have been shown to hold important leverage over the star formation \citep[e.g.,][]{2003MNRAS.341...54K, 2010ApJ...721..193P} and kinematic state \citep[e.g.,][]{2013ApJ...778L...2C, 2017ApJ...837...68C} of galaxies.
Given that in ours study we want to investigate the inter-dependence of a number of different parameters simultaneously, this can only be done using a sample of the size as that of the MaNGA survey.

This paper is structured as follows.
Section 2 presents an overview of data sources, sample selections and measurements.
Section 3 illustrates the interrelationship between star formation and kinematic state as seen from both $\mathrm{SFR} - \mathcal{M}_{\star}$ and $\lambda - \epsilon$ diagram and also the dependence on stellar mass, environment and visual morphology.
Section 4 discusses how to understand our results in the context of a $\Lambda \mathrm{CDM}$ universe.
Attempt of drawing implications for quenching mechanisms and comparison with relevant IFS studies are also made.
Lastly, Section 5 lists our main conclusions.

Throughout this paper we adopt a Chabrier IMF and standard values for cosmological parameters, $H_0=70\,km\,s^{-1}\,Mpc^{-1}$, $\Omega_m=0.3$ and $\Omega_{\Lambda}=0.7$, which are close to recent measurements \citep{2018arXiv180706209P}.


\section{Samples}\label{sec:sample}

The IFS data used in this study are taken from the SDSS Data Release 15 (DR15) \citep[SDSS15; ][]{2019ApJS..240...23A}.
DR15 comprises IFS data for 4597 unique galaxies observed via integral field units (IFUs) arranged in hexagon with effective diameters ranging from 12 to 32 arcsec \citep{2015AJ....149...77D}.
This corresponds to IFUs consisting of 19 to 127 fibres of 2 arcsec diameters that feed light into the two dual-channel BOSS spectrographs \citep{2013AJ....146...32S}.
The spectral range covers from 360 to 1030 nm with a median instrument broadening $\sigma_{\mathrm{inst}}\,\sim\,72\,\mathrm{km}\,s^{-1}$ \citep{2016AJ....152...83L} and a resulting typical spectral resolution $R \sim 2000$ \citep{2013AJ....146...32S, 2015AJ....150...19L}.

The selection scheme for galaxies that have been and will be observed by MaNGA is detailed in \citet{2017AJ....154...86W}.
Briefly, MaNGA aims at reaching a final IFS sample of $\sim$10,000 galaxies in the redshift range $0.01<z<0.15$.
Two of the three MaNGA subsamples, the Primary and Secondary sample which make up the majority the sample, are targeted to have spectroscopic coverage to 1.5 and 2.5 projected half-light radii ($R_{\rm e}$) respectively and are selected to have a flat number density distribution with respect to the $-$-band absolute magnitude $M_i$ (as a proxy for stellar mass).
Another subsample, the Colour-Enhanced sample, increases the number of galaxies in the low-density regions of colour–magnitude diagram by extending the redshift limits of the Primary sample in appropriate colour bins.
By such, the Primary plus the Colour-Enhanced sample make up a sample with a smoother coverage in colour-magnitude\footnote{Colour index: GALEX NUV minus SDSS i band. Magnitude: absolute magnitude in SDSS i band.}
diagram.
There are also ancillary targets included in DR15.
These targets are those of special interest but rare in a representative galaxy sample such as luminous AGNs and mergers.
Especially, no cuts are made on colour, morphology, or environment so that the galaxies observed by MaNGA are fully representative of the local galaxy population.

We adopt the same approach to data quality control as in \citet{2018MNRAS.477.4711G}.
This excludes IFS data a significant fraction of which are either flagged bad or show signs of being problematic.
It also excludes galaxies in a merger and too small galaxies as compared with the MaNGA beam size (See a detailed description in \citealt{2018MNRAS.477.4711G}).
In addition, as also recommended in \citet{2017AJ....154...86W}, galaxies in the ancillary sample are excluded as they reduce the sample representativeness of the local galaxy population.

Lastly, because of the limited spatial coverage, MaNGA data alone are not enough for measuring the total SFR (Fig. 2 in \citealt{2019ApJ...870...19G}).
Thus we cross match the sample with GALEX-SDSS-WISE Legacy Catalogue (GSWLC; references and more introduction in \autoref{subsec:gswlc}) for total SFR of MaNGA galaxies.
By doing this we only further put a limit that the sample is covered by the footprint of GALEX All-sky Imaging Survey (see \autoref{subsec:gswlc}), which overlaps with the footprint of SDSS spectroscopic survey for about 90 per cent.
So finally we reach a MaNGA IFS sample of 3279 galaxies, dubbed $\mathcal{S}_{\mathrm{MaNGA}}$.

To set up our "reference" of star formation for assessing the star formation level of galaxies, in a following section (\autoref{subsec:general}) we will define the SFMS by using local galaxy populations catalogued in GSWLC.
It includes $\sim$700 000 SDSS galaxies within the footprint of GALEX AIS with SDSS redshift below 0.3.
And this sample is dubbed $\mathcal{S}_{\mathrm{GSWLC}}$.
In such a way, through out this work there is only one source of SFR measurements so that the assessed star formation level of galaxies is self-consistent.


\section{Measurement sources}\label{sec:measure}

The key parameters in this work are stellar mass, SFR, the spin parameter $\lambda_{R_e}$ and ellipticity $\epsilon$ by which we quantify the star formation level and kinematic state of MaNGA galaxies.
In this section we concisely describe how these are measured while leaving the full-length description in the sources where we collected these measurements.
Together, we also describe the group identifications and visual morphological classifications used in this work.
They are respectively for studying the environmental dependence and giving a glance at the interrelationship between kinematic state and visual morphology.

\subsection{Ellipticity and the spin parameter in the half-light ellipse}\label{subsec:lame}

The measurements of ellipticity $\epsilon$ and the spin parameter are directly taken from \citet{2019arXiv191005139G} with the methods described in detail in \citet{2018MNRAS.477.4711G}.

\subsubsection{Determination of the half-light ellipse and ellipticity $\epsilon$}

The spin parameter $\lambda_{R_e}$ of each galaxy is measured within the half-light ellipse.
The ellipse is defined by ellipticity and area $A=\pi R_e^2$ where $R_e$ is the circular effective radius described in the following.

Firstly, the SDSS $r$-band photometry from NASA-Sloan Atlas\footnote{http://www.nsatlas.org}
\citep[NSA; ][]{2011AJ....142...31B} is fitted\footnote{This used the algorithm and \textsc{MgeFit} Python software package of \citet{2002MNRAS.333..400C} available at https://pypi.org/project/mgefit/} using the Multi-Gaussian Expansion (MGE) method \citep{1994A&A...285..723E, 2002MNRAS.333..400C}, which models the surface brightness via the sum of a certain number (12 is used for every galaxy in \citealt{2018MNRAS.477.4711G}) of two-dimensional Gaussians.

Then, the isophotal contour containing half the MGE total luminosity is determined by making use of the routine \textsc{mge\_half\_light\_isophote}\footnote{Included in the \textsc{JamPy} Python software package of \citet{2008MNRAS.390...71C} available at https://pypi.org/project/jampy/}
which implements steps (i) to (iv) found before equation 12 in \citet{2013MNRAS.432.1709C}.
And the ellipticity $\epsilon$ is calculated inside this half-light isophote as that of the inertia ellipse as \citep{2007MNRAS.379..418C}
\begin{equation}
(1-\epsilon)^2\,=\,q^{\prime 2}\,=\,\frac{\left< y^2 \right>}{\left< x^2 \right>}\,=\,\frac{\sum_{k=1}^{P} F_k y_k^2}{\sum_{k=1}^{P} F_k x_k^2}
\end{equation}
where $F_k$ is the flux of the $k$th pixel, with coordinates $(x_k,y_k)$ and the summation extends to the pixels inside the isophote.
The effective radius $R_e$ is defined such that a circle of radius $R_e$ covers the same area as the half-light isophote.
An empirical factor of 1.35 is further applied to $R_e$ in order to match the effective radius measurements from 2MASS \citep{2006AJ....131.1163S} and RC3 \citep{1991rc3..book.....D} combined (see Fig. 7 of \citealt{2013MNRAS.432.1709C}).

\subsubsection{The spin parameter $\lambda_{R_e}$}

In this work the spin parameter $\lambda_{R_e}$, a proxy for specific angular momentum (angular momentum per unit mass) of stars\footnote{Or more accurately, rotational velocity as a fraction of square root of velocity second moment weighted by flux and radius. Thus it indicates the relative importance of rotation in the overall motion and is similar to the spin parameter of halo \citep{2001ApJ...555..240B}.}
, is used as a diagnostic of galactic kinematic state.

To calculate $\lambda_{R_e}$, \citet{2019arXiv191005139G} took stellar kinematics from MAPS files which are the primary output of the Data Analysis Pipeline \citep[DAP; ][]{2019arXiv190100856W}.
Based on data products produced by the Data Reduction Pipeline \citep[DRP; ][]{2016AJ....152...83L}, the DAP applies the Penalised Pixel-Fitting method \citep[pPXF; ][]{2017MNRAS.466..798C} to extract the line-of-sight velocity distribution (LOSVD) by fitting a set of 49 families of stellar spectra from the MILES stellar library \citep{2006MNRAS.371..703S, 2011A&A...532A..95F} to the absorption-line spectra.
And then the data are spatially Voronoi binned \citep{2003MNRAS.342..345C} to achieve a minimum signal-to-noise ratio of $\sim10$ per spectral bin of width $70\,\mathrm{km}\,s^{-1}$ before the mean stellar velocity V and velocity dispersion $\sigma$ are extracted.

With stellar kinematics maps ready, $\lambda_{R_e}$ is calculated using equation 5 and 6 of \citet{2007MNRAS.379..401E}:
\begin{equation}
\lambda_{R}\,\equiv \, \frac{\left< R\,|V| \right>}{\left< R\, \sqrt{V^2+\sigma ^2} \right>}\,=\,\frac{\sum_{n=1}^{N} F_n R_n |V_n|}{\sum_{n=1}^{N} F_n R_n \sqrt{V_n^2+\sigma _n ^2}}
\end{equation}
with the summation performed over N pixels within the radius R (for $\lambda_{R_e}$ within the half-light ellipse determined before) and $F_n$, $V_n$, $\sigma _n$, being the flux, mean velocity and velocity dispersion of the nth pixel respectively.
By incorporating radial distance of pixels in addition to flux, $\lambda_{R_e}$ gives less weight compared with $V/\sigma$ to central pixels which usually have nearly zero velocity and high velocity dispersion.
And this can also make $\lambda_{R_e}$ sensitive to some spatial structure like rings that locate at larger radii.

The MaNGA beam size is sometimes non-negligible compared with the area of the half-light ellipse, and this tends to smear out the LOSVD and make the observed value of $\lambda_{R_e}$ lower than the intrinsic one.
To account for this \citet{2019arXiv191005139G} includes an analytic correction to observed $\lambda_{R_e}$ which is derived in \citet{2018MNRAS.477.4711G}.
Briefly, \citet{2018MNRAS.477.4711G} quantified the effect of atmospheric smearing by measuring $\lambda_{R_e}$ of galaxy kinematic models \citep[JAM; ][]{2008MNRAS.390...71C} convolved with a Gaussian PSF for a range of PSF sizes.
And this method of correction has been tested in \citet{2019MNRAS.483..249H} using realistic and independent models and shown to do a good job in recovering the intrinsic $\lambda_{R_e}$ with no systematic deviations for a range of PSF widths.

\subsection{Star formation rate and stellar mass}\label{subsec:gswlc}

Stellar mass and star formation rate of the MaNGA galaxies are taken from the version X2 of GALEX-SDSS-WISE Legacy Catalogue
\footnote{http://pages.iu.edu/\textasciitilde salims/gswlc/}
\citep[GSWLC-X2,][]{2016ApJS..227....2S,2018ApJ...859...11S}.
It is a value-added catalogue for SDSS galaxies in the redshift range $0.01<z<0.3$ within GALEX All-sky Imaging survey footprint \citep{2005ApJ...619L...1M}.
Using the state-of-the-art spectral energy distribution (SED) modelling technique \citep[CIGALE; ][]{2009A&A...507.1793N}, stellar mass and star formation rate are derived by fitting the SED of galaxies consisting of two GALEX UV bands, five SDSS optical plus near infrared bands and also one mid infrared band (22 microns or 12 microns when the former is not available) from WISE \citep{2010AJ....140.1868W}.
In the footprint of GALEX All-sky Imaging survey, there are also GALEX Medium Imaging Survey and Deep Imaging Survey nested with progressively longer exposure time.
And GSWLC-X2 has utilised the deepest available UV image for each galaxy from these sources.

\begin{figure*}
    \includegraphics[width=0.49\textwidth]{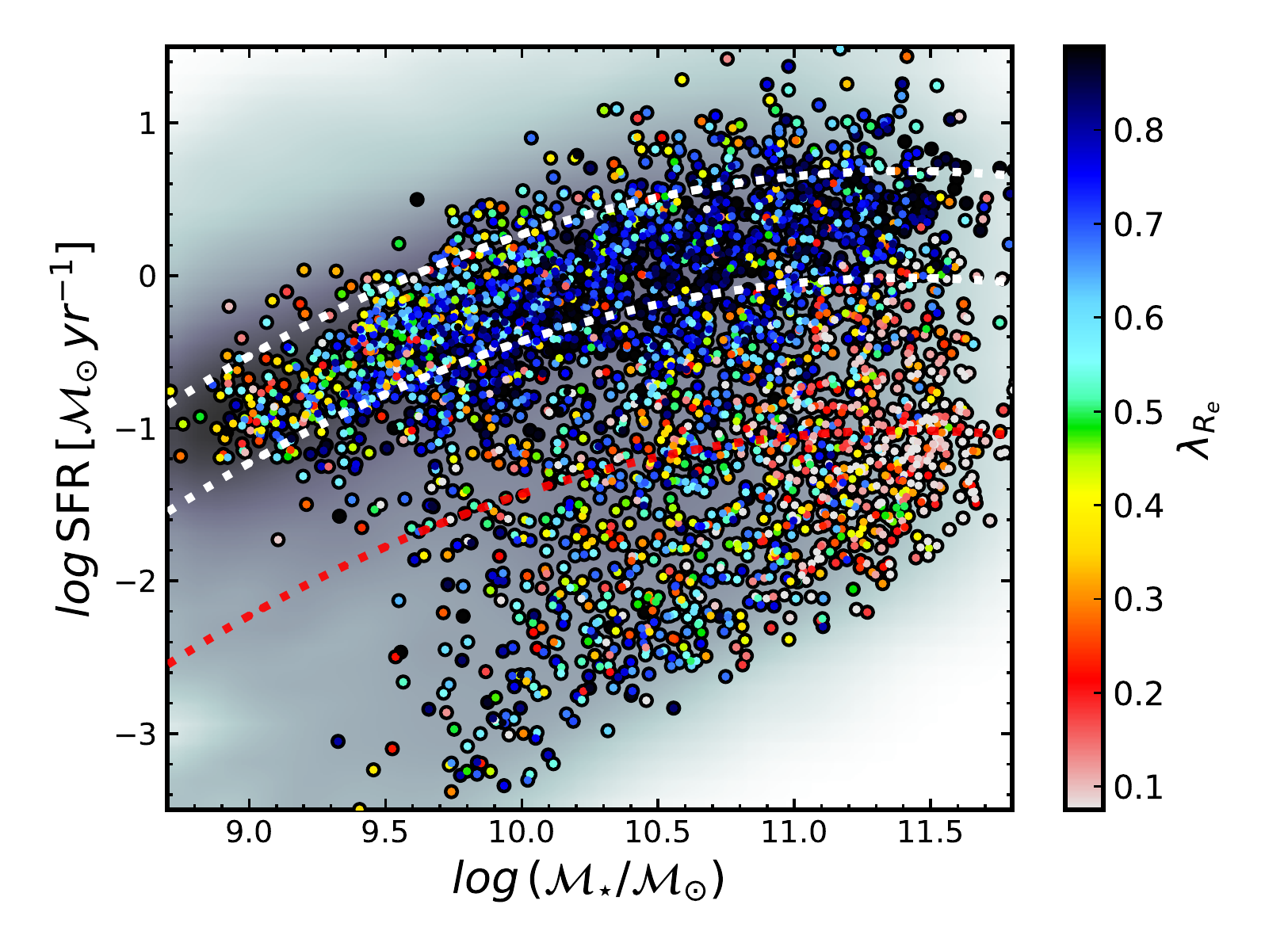}
    \includegraphics[width=0.49\textwidth]{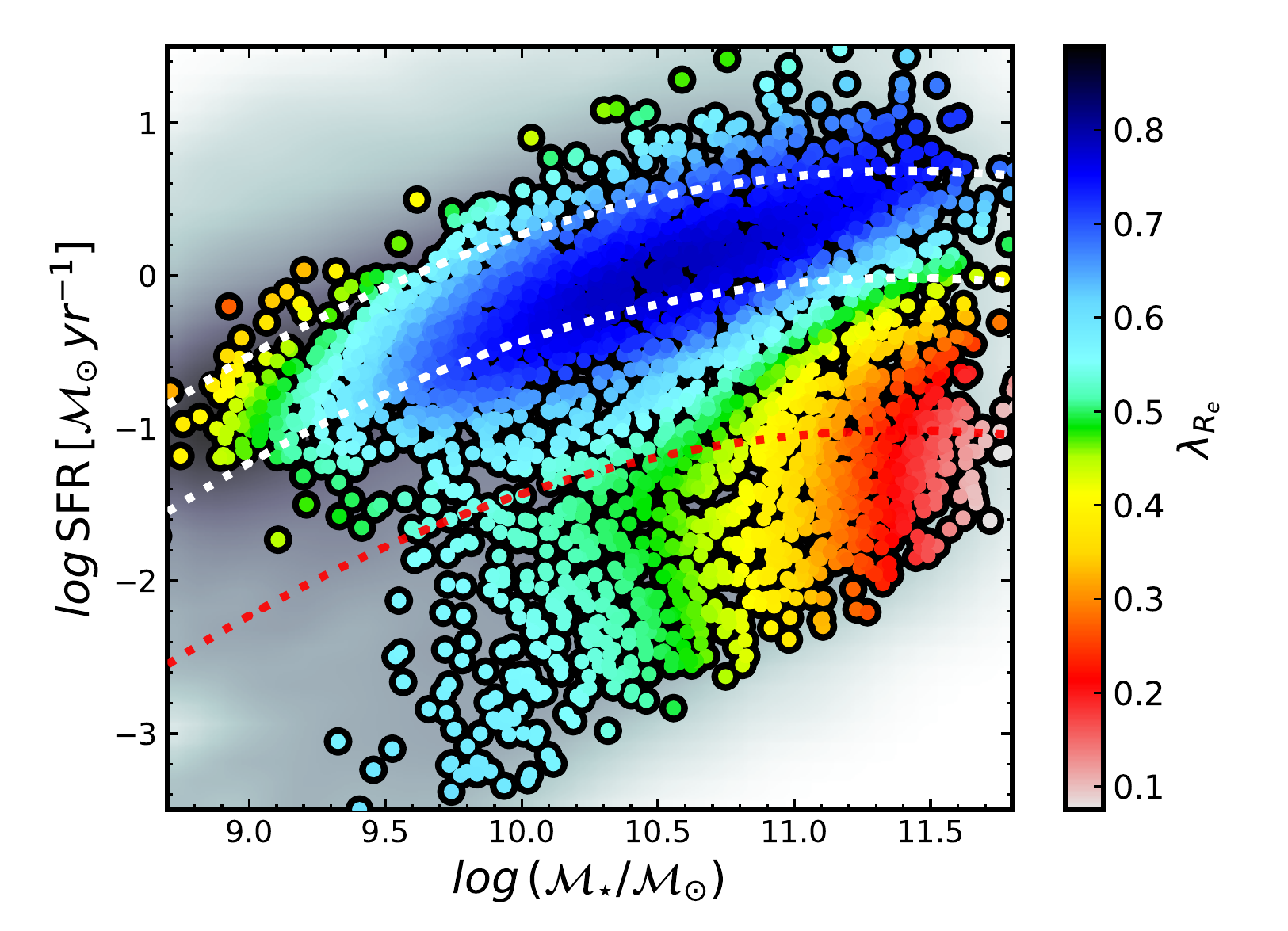}
    \caption{Left: Projection of the spin parameter $\lambda_{R_e}$ onto $\mathrm{SFR}-\mathcal{M}_{\star}$ plane for galaxies in sample $\mathcal{S}_{\mathrm{MaNGA}}$.
    Background grey scale shows the volume corrected probability density of galaxies in $\mathcal{S}_{\mathrm{GSWLC}}$ derived from kernel density estimation.
    White dotted lines mark $\pm 0.35 $ dex from the star formation main sequence, the ridge line of the background density distribution, within which galaxies are defined star-forming galaxies.
    Red dotted line is one dex below the lower white dotted line and galaxies beneath are considered passive/quenched galaxies.
    Right: LOESS smoothed version of the left (see more introduction in the text).
     }
    \label{fig:sfrm}
\end{figure*}


\subsection{Environment indicators}\label{subsec:env}

The complex nature of galaxy environment makes any one kind of environment definition insufficient as a thorough description of it.
So in this work we will show the dependence of kinematic morphology-star formation relation on several types of environment indicators:
\begin{itemize}
  \item[--] $\mathcal{N}$: Group richness, i.e. number of member galaxies in a group.
  \item[--] $\Sigma_3\,(\Sigma_{10})$: Projected galaxy number density within a circle of radius equal to the distance to the third (tenth) nearest neighbour.
  \item[--] Group identity: Central galaxy or satellite, depending on if the galaxy is the most massive galaxy in the group (central) or not (satellite).
\end{itemize}

Group richness generally correlates with the total luminous mass of a group and thus also with the total dark mastter mass given the galaxy-halo connection \citep{2018ARA&A..56..435W}.
So a large $\mathcal{N}$ usually means a massive galaxy cluster with a massive dark matter halo.
In such a deep potential, strong tidal field as well as hot intracluster medium are able to significantly alter galaxy properties gravitationally and hydrodynamically (see a review in \citet{2006PASP..118..517B}).
And the $\mathcal{N}$ used in this work is from a group catalogue for MaNGA galaxies constructed in \citet{2019arXiv191005136G} using the local group finder \textit{TD-ENCLOSER} presented in \citet{2019arXiv191005135G}.
\textit{TD-ENCLOSER} is based on the kernel density estimation (KDE) method and is optimised for obtaining the local galaxy environment.
Descending from peaks in density filed, the algorithm assigns galaxies to these peaks so that galaxy groups "grow" around these peaks.
A specified hard threshold is set to exclude outliers in underdense regions and at the group edges outliers are clipped below a soft (blurred) interior density level.
In \citet{2019arXiv191005136G} \textit{TD-ENCLOSER} is applied for each MaNGA galaxy in a cylinder centered on the galaxy with hight ranging from $600\,\mathrm{km}\,s^{-1}$ by default up to $6000\,\mathrm{km}\,s^{-1}$ for large clusters.
And galaxies falling in the cylinder are out of a sample obtained by combining SDSS spectroscopic and photometric catalogues so that incompleteness in the SDSS spectroscopic catalogue (e.g. due to fiber collision) is accounted for.

The second sort of environment indicator we use is the galaxy surface number density measured within a circle of radius the distance to the nth nearest neighbour, which has been extensively chosen as gauge of environment in the literature \citep[e.g., ][]{1980ApJ...236..351D, 2006MNRAS.373..469B, 2009MNRAS.393.1324B, 2012ApJ...757....4P}.
Specifically, in this work we make use of density based on 3rd ($\Sigma_3$) and 10th ($\Sigma_{10}$) nearest neighbour provided by \citet{2019arXiv191005136G} defined as $\Sigma_n=N/(\pi D_n^2)$ with $D_n$ in Mpc and calculation performed in a cylinder centered on the galaxies of hight $600\,\mathrm{km}\,s^{-1}$ ($\Delta V=\pm 300\,\mathrm{km}\,s^{-1}$).
$\Sigma_3$ has been found to give a very smooth kinematic morphology-density relation in \citet{2011MNRAS.416.1680C}, suggesting it as a close proxy for the regulator of the relation.
The same study has also shown that compared with $\Sigma_3$, $\Sigma_{10}$ is a better indicator of environment on a larger scale so that it can clearly separate field and cluster galaxies.

Lastly, we also adopt the central/satellite dichotomy defined by a group catalogue constructed for SDSS galaxies in \citet{2012ApJ...752...41Y}.
This group catalogue is probably the most utilised one among studies of SDSS galaxies, and a large range of SDSS galaxy properties bifurcate after galaxies are differentiated under this dichotomy \citep[e.g., ][]{2013MNRAS.436...34C, 2014MNRAS.438..262P, 2017MNRAS.464.1077W}.
Instead of using a kernel-based finder, \citet{2012ApJ...752...41Y} applied an iterative group finder based on friends-of-friends algorithm, which is an updated version of that in \citet{2005MNRAS.356.1293Y}.
In short, halo properties inferred from tentative galaxy groups (identified using friends-of-friends method) are further used to update group membership of galaxies, which in turn updates halo properties.
\citet{2012ApJ...752...41Y} applied this algorithm to three galaxy samples with slightly increasing galaxy completeness but decreasing redshift reliability.
Here we choose the "PetroB" version for a trade-off between the two.
And more than 95\% of MaNGA galaxies in $\mathcal{S}_{\mathrm{MaNGA}}$ have group information therein.


\begin{figure*}

    \includegraphics[width=0.98\textwidth,trim={0 1cm 0 0},clip]{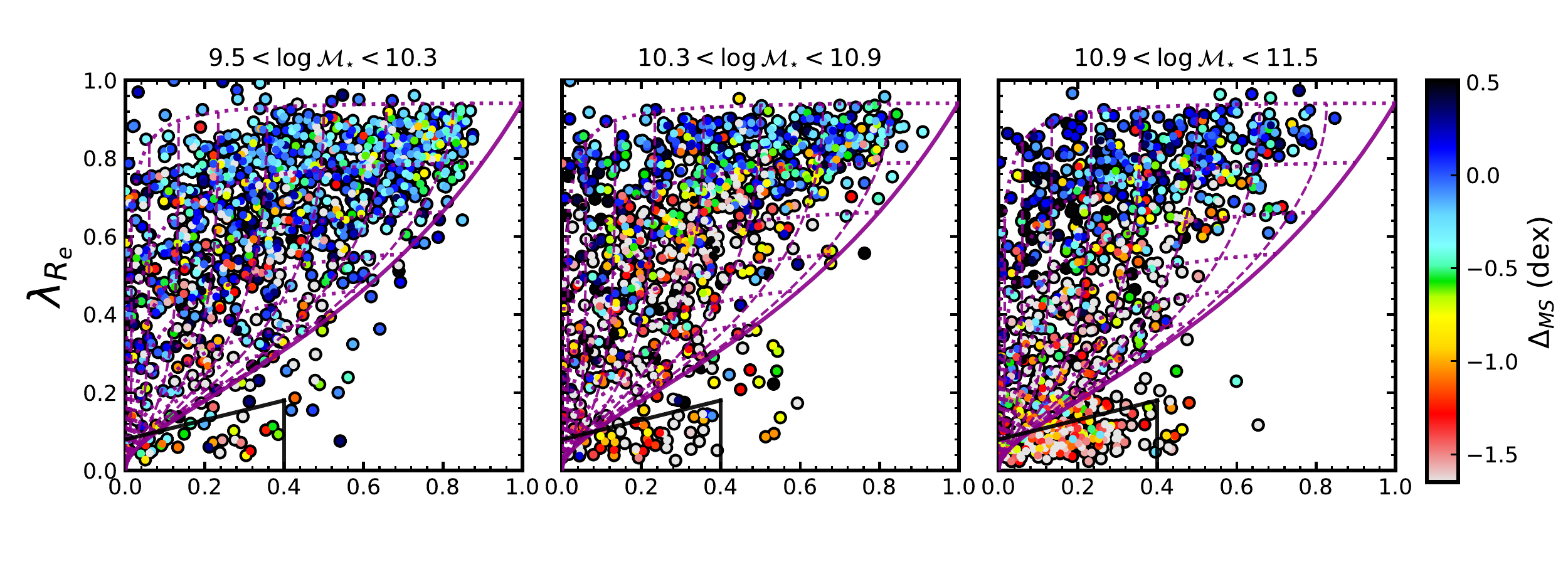}
  \includegraphics[width=0.98\textwidth,trim={0 0 0 1cm},clip]{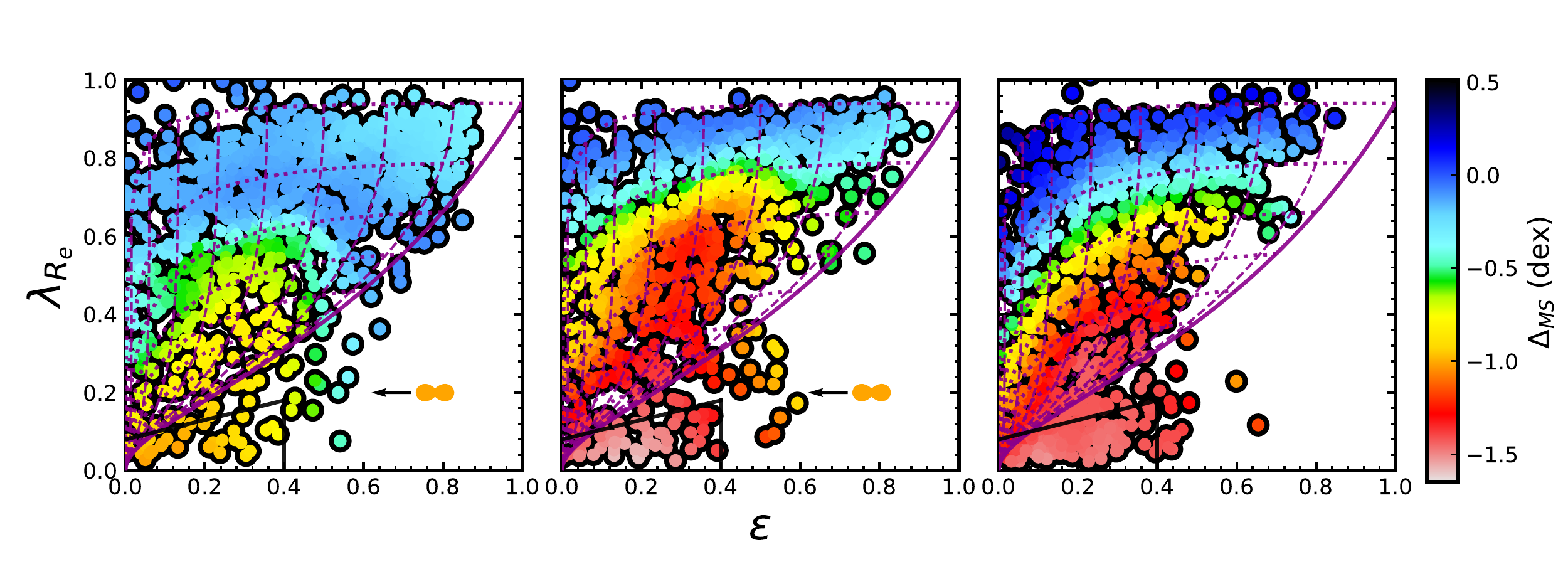}
    \caption{Upper: Projection of $\Delta _{\mathrm{MS}}=\mathrm{SFR}-\mathrm{SFR}_{\mathrm{MS}}(\mathcal{M}_{\star})$ onto $\lambda_{R_e} - \epsilon$ plane.
    The sample is split into three mass bins in which the data have good coverage across $\mathrm{SFR}-\mathcal{M}_{\star}$ plane.
    In each panel, theoretical predictions based on tensor virial theorem for oblate rotators following a certain anisotropy relation $\beta_z =0.7 \times \epsilon _{\mathrm{intr}}$ are shown: The solid magenta line corresponds to a edge-on view of the theoretical oblate rotators with different intrinsic ellipticity.
    The dotted magenta lines are predictions for different fixed intrinsic ellipticity (zero to one as going above) seen at varying inclination (face-on to edge-on from left to right).
    And dashed magenta lines are predictions for the other way around, i.e. fixed inclination but varying intrinsic ellipticity.
    The region confined by the black solid lines denotes where favoured by slow rotators.
    Lower: LOESS smoothed version the upper row.
    The orange peanut symbols denote galaxies with counter-rotating disk components which feature two peaks away from center on velocity dispersion maps, i.e. two-sigma galaxies.
    And the arrows point to the region populated by two-sigma galaxies.
    Note their higher star formation level compared with galaxies of similar $\lambda_{R_e}$ at lower $\epsilon$.
    }
    \label{fig:le}
\end{figure*}


\subsection{Galaxy Zoo visual morphological classifications}\label{subsec:gz}

For the purpose of comparing galaxy kinematic state with morphology, we take Galaxy Zoo 1 (GZ1) visual morphological classifications of SDSS galaxies from the Galaxy Zoo citizen project \citep{2008MNRAS.389.1179L, 2011MNRAS.410..166L}.
Galaxies catalogued in GZ1 are classified into three types - elliptical, spiral and indeterminate galaxies - determined by weighted mean of votes with a threshold of 80 per cent.
A full description of the weighting scheme and the debiasing method accounting for effects of size, luminosity, distance etc. are provided in \citet{2008MNRAS.389.1179L} and \citet{2009MNRAS.393.1324B} respectively.

It is important to note that the GZ1 classification scheme has, by design, some important differences with respect to the popular Hubble classification which has been used by astronomers for a century \citep{Hubble1926,Sandage1961,deVaucouleurs1959}.
In fact, spiral galaxies classified in GZ1 are defined as heaving evidence of disks, but not necessarily spiral arms. This means that most S0 galaxies will be included by GZ1 into the spiral galaxy category.
So in the following we directly refer to them as GZ1 ellipticals (GZ1-E) and GZ1 spirals (GZ1-S).
As for the indeterminate type in GZ1 (GZ1-I), they are galaxies whose morphology the project volunteers do not know for sure.
This includes those with too small apparent sizes or irregular morphology for example, and in many cases, composite bulge-disc systems in which neither the bulge nor disc clearly dominates \citep{2014MNRAS.440..889S}.

And, there are more than 95\% of galaxies in sample $\mathcal{S}_{\mathrm{MaNGA}}$ catalogued in GZ1.


\section{Results}

\subsection{The mass dependent kinematic-morphology vs star-formation relation}\label{subsec:general}

With this unprecedentedly large sample of galaxies with IFS data, now we are able to attain a thorough view of the kinematic state of galaxies on $\mathrm{SFR}-\mathcal{M}_{\star}$ diagram, down to a mass of $\sim 10^{9.5} \mathcal{M}_{\odot}$.
The results are shown in \autoref{fig:sfrm}.

Serving as a star formation benchmark, we have defined the so-called "star formation main sequence" (SFMS) as the ridge line of the probability density distribution of galaxies in $\mathcal{S}_{\mathrm{GSWLC}}$ in redshift range $0.01<z<0.08$ (see \aref{app:sfms} for more details).
The upper limit of our redshift range was adopted by \citet{2019arXiv191005136G} to have a more reliable kinematic classification and higher redshift completeness for the environmental estimate. But this makes our SFMS potentially not representative for MaNGA galaxies with $\mathcal{M}_{\star}\gtrsim10^{11} \mathcal{M}_{\odot}$.
This is because under the MaNGA sample selection scheme \citep{2017AJ....154...86W}, a limit in redshift $z<0.08$ implies a luminosity limit $M_i\gtrsim-23.0$ and results in a drop in the fraction of the most luminous galaxies with respect to the full MaNGA sample. However, we have checked that our SFMS matches exactly the ridge line of distribution of these massive galaxies at $z>0.08$.
The two white dotted lines in the two panels mark the position 0.35 dex above/below SFMS and they roughly define $\pm \sigma$ scatter which is consistent with results in the literature (refer to a census in \citealt{2014ApJS..214...15S}), and also with the standard deviation of fitted Gaussians in stellar mass bin.
We define the galaxies within this band the star-forming galaxies hereafter and those above it the star-bursting galaxies.
While the red dotted line lies at one dex below the lower white dotted line so that galaxies below the red have SFR at least one dex lower than star-forming galaxies of the same stellar mass and we define these galaxies the passive galaxies.
We define the galaxies in between the green valley galaxies.

In the left panel of \autoref{fig:sfrm}, $\lambda_{R_e}$ of about 3,200 MaNGA galaxies in $\mathcal{S}_{\mathrm{MaNGA}}$ is mapped onto $\mathrm{SFR}-\mathcal{M}_{\star}$ plane.
And in the right panel, we smooth the data using locally weighted regression method
LOESS by \citet{Cleveland1988} as implemented\footnote{We used the Python package \textsc{loess} v2.0.11 available from https://pypi.org/project/loess/} by \citet{2013MNRAS.432.1862C}. LOESS is designed to uncover underlying mean trends by reducing observational errors and intrinsic scatter. Statistically, LOESS tries to estimate what one would infer by simply averaging values in small bins, if the sample was much larger than the present one.
In application of LOESS smoothing, we adopt a smoothing factor \texttt{frac} = 0.3, and a linear local approximation.
Different scales of vertical and horizontal axes are accounted for by rotating and re-normalizing the coordinates so that the ellipse of inertia of the galaxy distribution reduces to a circle in every projection.

LOESS reveals that star-forming galaxies on the SFMS are generally dynamically cold, except for very low-mass ones ($\mathcal{M}_{\star} \sim 10^{9} \mathcal{M}_{\odot}$) whose $\lambda_{R_e}$ on average is comparable with some massive passive galaxies. The increase in $\lambda_{R_e}$ at the lowest mass may be related to the increased roundness of dwarf spheroidal (Sph) galaxies illustrated in Fig.~24 of \citet{2016ARA&A..54..597C} which is generally attributed to tidal disturbances, as reviewed by \citet{2012ApJS..198....2K}. This may also be related to the signal of a decrease of cold orbit fraction at low mass end reported in \citet{2018NatAs...2..233Z}.

Above the SFMS we see a hint of $\lambda_{R_e}$ drop.
A possible explanation for this drop may be an inclination bias, i.e. galaxies being predominantly face-on making the SFR overestimated (less dust attenuation) and $\lambda_{R_e}$ lower than a randomly oriented sample.
In that case, these star-bursting galaxies can actually be just more face-on version of the population on the SFMS.
But, a visual check of the SDSS g-r-i composite images indicates some intrinsic differences and in particular a higher incidence of irregular features and blue star-forming clumps among these star-bursting galaxies, as compared with a similar number of galaxies on the SFMS.
This together with another argument based on the distributions on $(\lambda-\epsilon)$ diagram (\aref{app:inc}) to a large extent rule out the above possibility.
And we note that this drop of $\lambda_{R_e}$ seems to be in line with an increase of \citet{Sersic1968} index \citep{2011ApJ...742...96W} and bulge to total ratio \citep{2017A&A...597A..97M} found by works based on light distribution.

Among galaxies below the SFMS, the mass trend is reversed.
$\lambda_{R_e}$ of these galaxies decreases with mass with the trend especially sharp at the massive end.
The red region at mass above $\sim 10^{11.3} \mathcal{M}_{\odot}$ is consistent with the previously found characteristic mass where slow rotators starts dominating \citep{2011MNRAS.414..888E, 2013ApJ...778L...2C, 2016ARA&A..54..597C}.
Vertically, namely viewing T-C relation from a kinematic perspective, there also exists an apparent mass trend.
Consistent with what we expect from the original T-C relation, $\lambda_{R_e}$ of massive galaxies, especially the most massive ones, reduces drastically along with their decreasing SFR.
However at mass below $10^{10} \mathcal{M}_{\odot}$, on average $\lambda_{R_e}$ only reduces by about 0.1 when SFR changes more than two orders of magnitude.
The same pattern can also be seen in the original data on the left, although trends are more difficult to see due to the significant intrinsic scatter hiding the underlying mean trend.
In \aref{app:lowspinsf} we compare the kinematics of low-mass and low-spin galaxies to the kinematics of massive low-spin ones. We conclude that those with lower mass generally indeed have complex velocity field while the low spin of the more massive systems are to a significant extent due to their low inclination.

Note that for individual galaxies, without the knowledge of their projected ellipticity, comparing alone their $\lambda_{R_e}$ would be meaningless given the obvious variation of $\lambda_{R_e}$ with the galaxies inclination.
However here, given the sample size as well as the fact that the selection scheme for MaNGA galaxies does not bias them for certain inclinations, the above results are meaningful and are indeed confirmed when taking ellipticity into account in the $(\lambda_{R_e},\epsilon)$ diagram (\autoref{fig:env2}).

We note that the PSF to galaxy angular size ratio of low-mass passive galaxies are generally larger and this means larger smearing correction to them.
However, this is only a small effect.
On average, because of dedicated sample selection and IFU allocation, MaNGA has achieved a roughly similar PSF (full width at half maximum) to galaxy (in terms major axis of half light ellipse) angular size ratio on $\mathrm{SFR}-\mathcal{M}_{\star}$ plane at the value $\sim\, 0.25$.
This is only not the case for part of region at low mass and low SFR region where the ratio can be $\sim\, 0.4$.
And when translated to its effect on $\lambda_{R_e}$, the smearing correction can result in only $\sim\, 0.05$ difference due to different PSF to galaxy angular size ratio between low mass star-forming and passive galaxies.

\autoref{fig:sfrm} shows $\lambda_{R_e}$ as a function of position on $\mathrm{SFR}-\mathcal{M}_{\star}$ plane while a complementary view is provided by \autoref{fig:le}.
The distance to SFMS ($\Delta _{\mathrm{MS}}=\mathrm{SFR}-\mathrm{SFR}_{\mathrm{MS}}(\mathcal{M}_{\star})$ ) of galaxies are projected onto $\lambda - \epsilon$ plane with galaxies divided into three mass bins and again the original data (upper row) and LOESS smoothed ones (lower row) are shown.
The superposed magenta line is the anisotropy-shape boundary $\beta_z =0.7 \times \epsilon _{\mathrm{intr}}$ of \citet{2007MNRAS.379..418C}, where $\beta _z=1-\sigma _z ^2\,/\,\sigma _R ^2$ quantifies the vertical to radial velocity dispersion ratio while $\epsilon _{\mathrm{intr}}$ is intrinsic ellipticity, i.e. ellipticity seen edge-on.

This relation was first projected onto the $(V/\sigma, \epsilon)$ diagram using the formalism of \citep{2005MNRAS.363..937B} based on tensor virial theorem. Then it was transformed into the corresponding relation for $(\lambda_{R_e}, \epsilon)$ using the empirical calibration of \citep{2007MNRAS.379..401E, 2011MNRAS.414..888E}.
The solid magenta line corresponds to an edge-on view of the prediction with varying intrinsic ellipticity.
While the dotted and dashed lines are predictions with fixed intrinsic ellipticity but varying inclination and fixed inclination but varying intrinsic ellipticity respectively.
The fact that the distribution of regular rotators can be well described by the solid magenta line and its projection at different inclinations, their lack kinematic misalignment, combined with results from dynamical models, indicates these galaxies are a family of galaxies with disks seen at different inclinations \citep[see review by][]{2016ARA&A..54..597C}.
In addition, the black solid lines also shown define the region occupied by slow rotators (equation 19 in \citealt{2016ARA&A..54..597C}).

LOESS smoothed data show that in the most massive bin, the galaxies with a certain star formation levels follow closely the predicted tracks for oblate rotators of fixed intrinsic ellipticity, in fact the colour contours crudely follow the dotted lines in the $(\lambda_{R_e}, \epsilon)$ diagram.
This corresponds to the drastic $\lambda_{R_e}$ reduction with decreasing SFR among massive galaxies that is shown in \autoref{fig:sfrm}.
And both of them are reflection of the close relationship between stellar populations and kinematic state of massive galaxies.
Toward lower mass, the stratification pattern becomes less clear which means the relationship is worse.
This results from the fact that can be seen in \autoref{fig:sfrm} that a growing number of galaxies with high $\lambda_{R_e}$ reach low SFR region while many with low $\lambda_{R_e}$ appear on the SFMS.
Such behaviour makes the relation weaker and smears out the stratification pattern we see for galaxies in the most massive bin.


\begin{figure*}
  \includegraphics[width=0.98\textwidth,trim={0 0.6cm 0 0},clip]{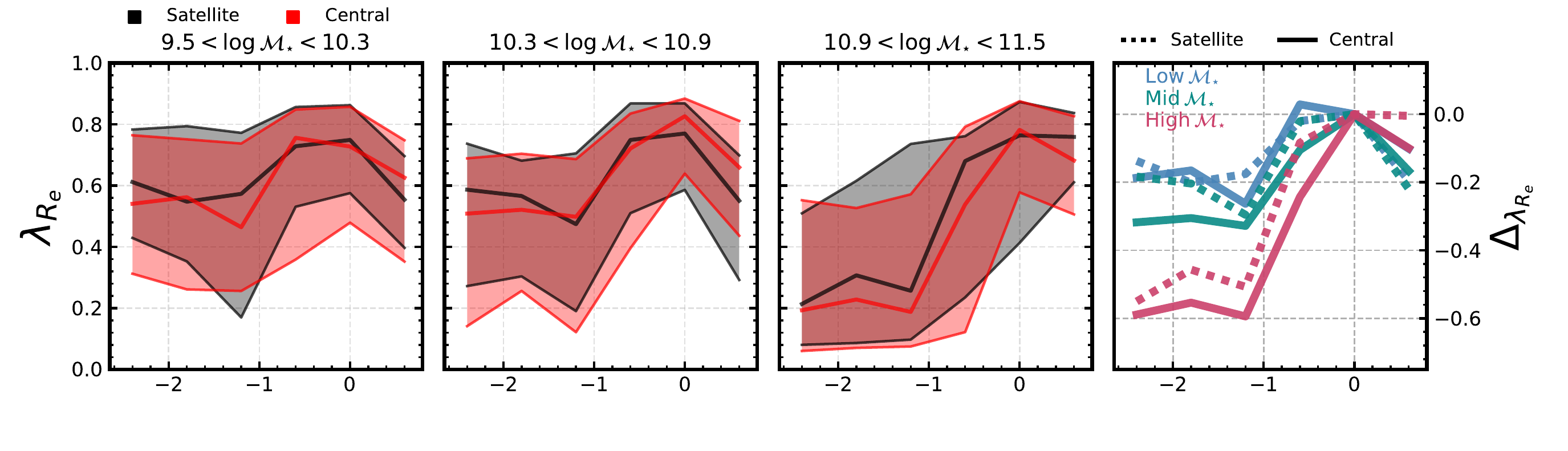}
    \includegraphics[width=0.98\textwidth,trim={0 0.6cm 0 0},clip]{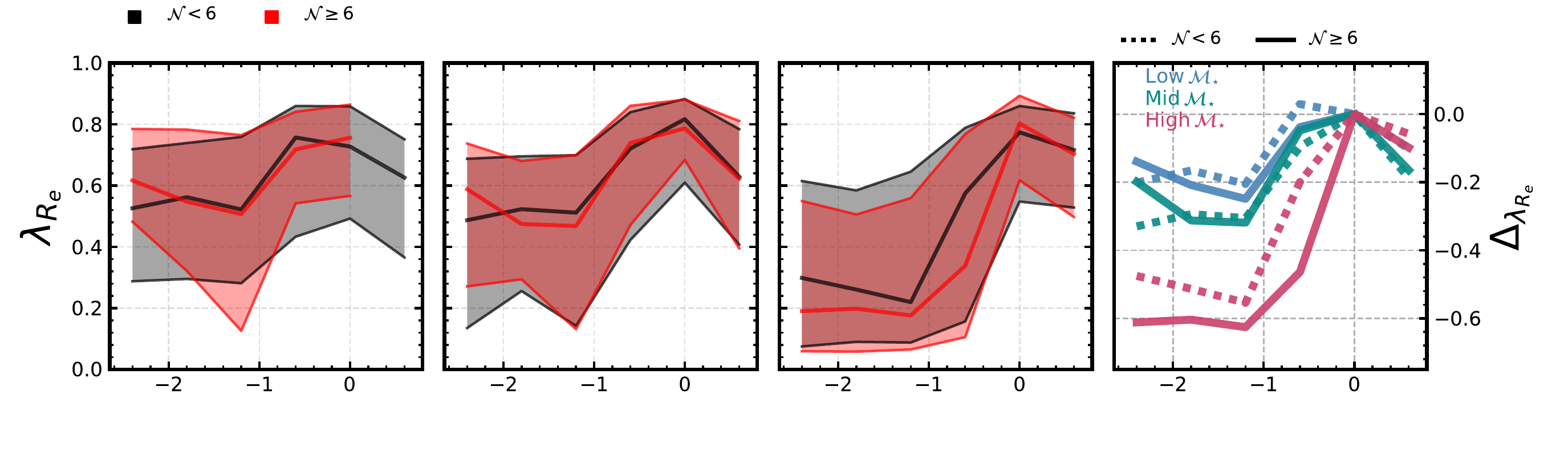}
  \includegraphics[width=0.98\textwidth,trim={0 0.6cm 0 0},clip]{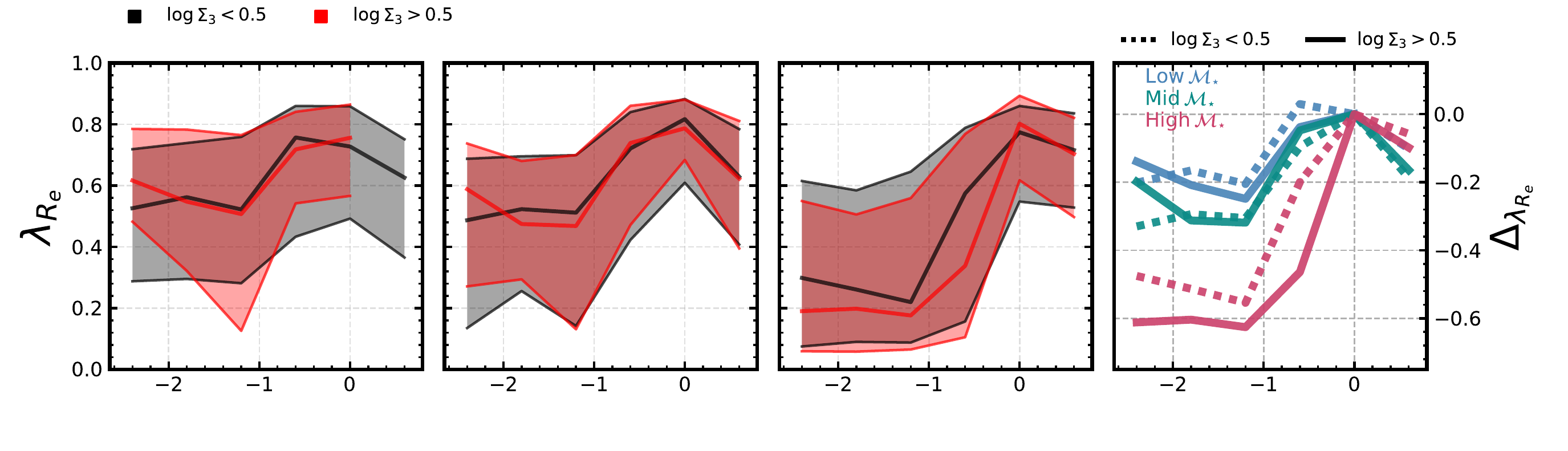}
  \includegraphics[width=0.98\textwidth]{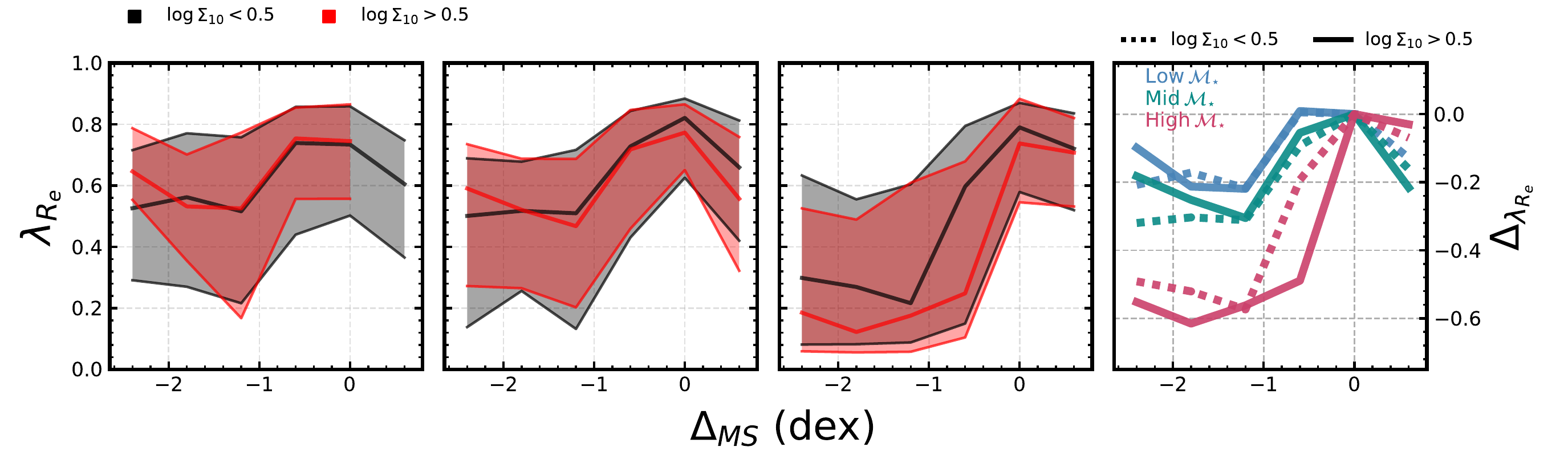}
    \caption{Environmental dependence of $\lambda_{R_e}$ as a function of $\Delta _{\mathrm{MS}}$ in three stellar mass bins.
    From the first row downward shows the dependence on classification in central/satellite dichotomy, group richness, galaxy surface number density calculated in a circle defined by the distance to the third and tenth nearest neighbours respectively ($\Sigma_3\, \& \, \Sigma_{10} $).
    For the three columns on the left, in each panel the median $\lambda_{R_e}$ and the band including inner 68\% of data points are displayed.
    Only $\Delta _{\mathrm{MS}}$ bins of 10 or more galaxies are shown.
    But all bins except the bin of star-bursting galaxies usually have data points much more than 10.
    In the last column, the median relations in certain mass and environment bins normalised to their own values on SFMS are compared, summarising the results on the left.
    }
    \label{fig:env1}
\end{figure*}


\begin{figure}
  \includegraphics[width=0.47\textwidth,trim={0 3.5cm 0 0},clip]{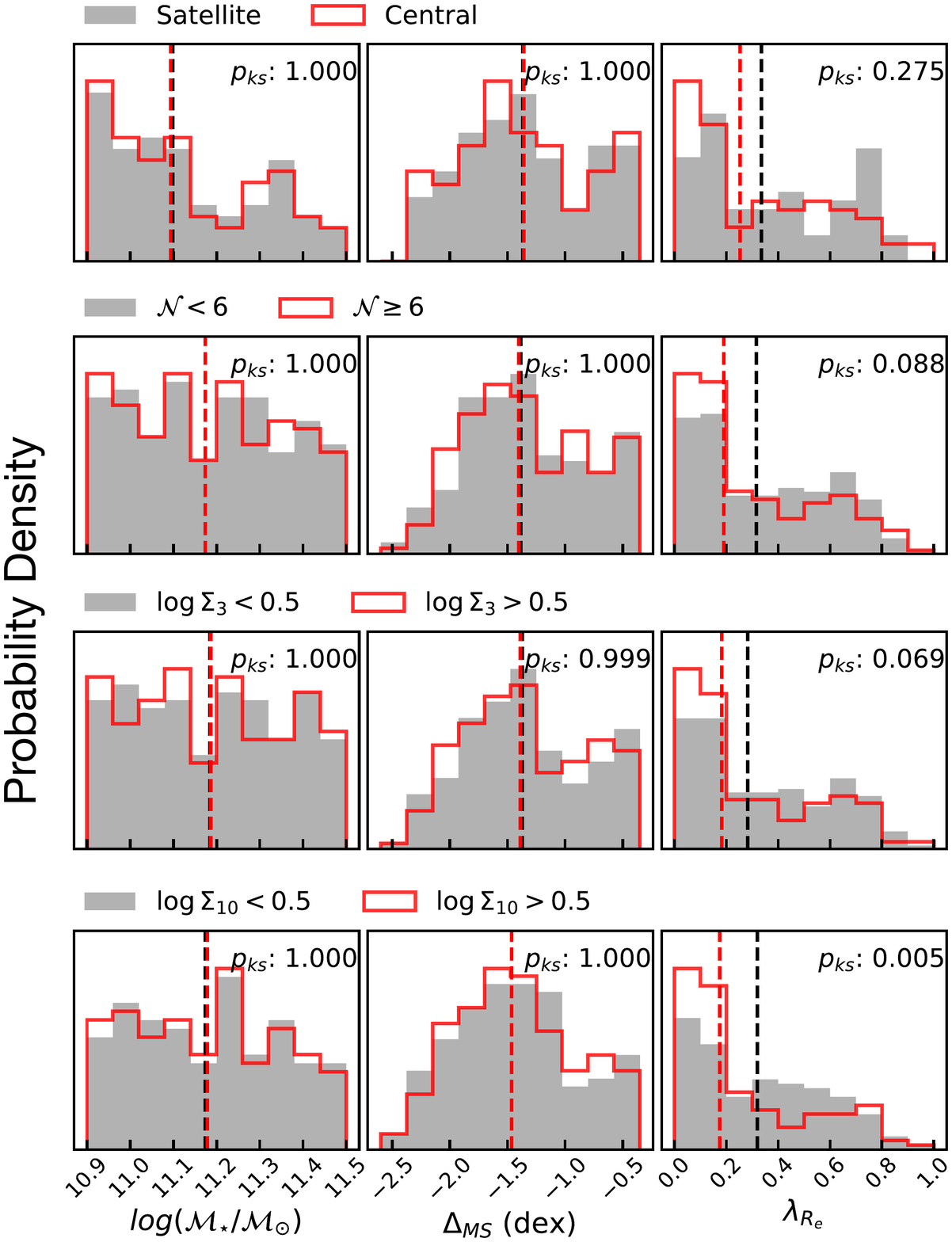}
    \caption{Distributions of $\mathcal{M}_{\star}$ (left column) and $\Delta_\mathrm{MS}$ (middle column) and $\lambda_{R_e}$ (right column) for massive galaxies below the SFMS in different environment (four indicators in the same row order as in \autoref{fig:env1}).
    In each row, the two samples are matched in $\mathrm{log}\, \mathcal{M}_{\star}$ and $\Delta_\mathrm{MS}$ with tolerance 0.02 and 0.1 dex respectively.
    In each panel median values are marked by dashed lines and the p-value of KS test is also shown.
    }
    \label{fig:kstest}
\end{figure}


\begin{figure*}

    \includegraphics[width=0.98\textwidth,trim={0 1cm 0 0},clip]{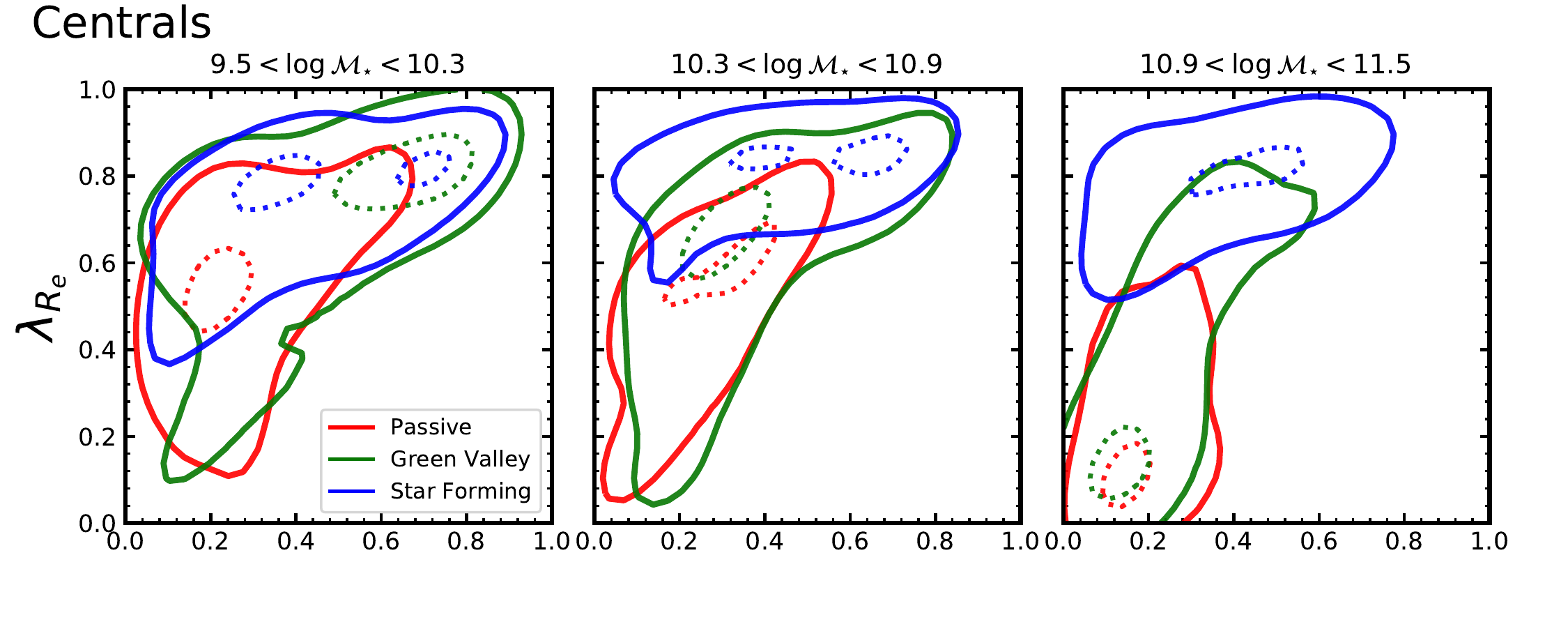}
  \includegraphics[width=0.98\textwidth]{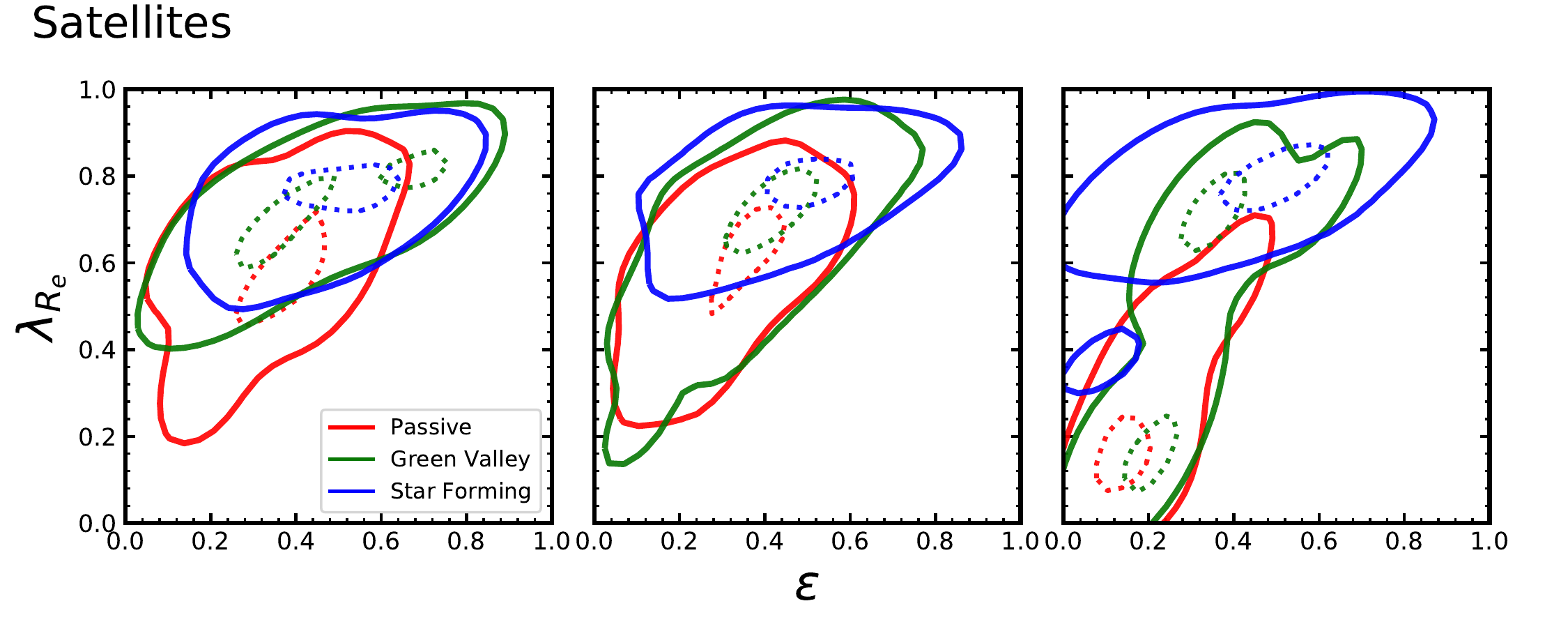}
    \caption{Confirming the result in the first row of \autoref{fig:env1} by further taking ellipticity into account.
    The Upper and lower row are for central and satellite galaxies respectively.
    Each panel shows in a certain stellar mass bin the probability density distributions of galaxies with certain star formation levels (star-forming, green valley and passive) derived via kernel density estimation.
    Solid lines mark the contours enclosing 68 per cent of total probability illustrating the position of bulk distribution, and the contours enclosing 10 percent are denoted as dotted lines showing the position of peaks.
    }
    \label{fig:env2}
\end{figure*}


It is interesting to point out two features in the diagram.
There is a group of galaxies mainly visible for the two low mass bins with unexpectedly low $\lambda_{R_e}$ for their star formation levels.
In the left two panels, they start to appear at the high ellipticity tip of populations of certain star formation levels making the tracks bending toward low $\lambda_{R_e}$ region.
Among these "outliers" the most outstanding ones are located in the box $0<\lambda_{R_e}<0.3\ \& \ 0.4<\epsilon<0.6$.
Thus we inspected the kinematic maps of galaxies in the two low mass bins in this region with a significantly larger star formation ($\Delta _{\mathrm{MS}}>-1$) with respect to rounder galaxies with the same $\lambda_{R_e}$.
There are 12 of these galaxies and their maps are shown in the \aref{app:straggler}.
Strikingly, at least 7 of them are "two-sigma" galaxies \citep{2011MNRAS.414.2923K}, with two symmetric peaks of velocity dispersion away from the centre, which indicate that they are actually galaxies with two counterrotating stellar disk components (see Fig. 12 of \citealt{2016ARA&A..54..597C}).
The large star formation is due to the fact that "two-sigma" galaxies form a physically homogeneous class with the rest of the fast rotators ETGs (Sec.~3.5.3 of \citealt{2016ARA&A..54..597C}) and this explains why they also have similar star formation properties as their non-counterrotating counterparts. In fact we can argue that the star formation provides a way to recognize counterrotating disks on the $(\lambda_{R_e}, \epsilon)$ diagram, even when the spatial resolution does not allow one to directly recognize the "two-sigma" kinematic feature.
The fact that not many massive galaxies show this feature shows that counterrotating disks are favoured in less massive systems.
And we only see them at the high ellipticity tips does not mean these counterrotating galaxies are only expected to have high ellipticity.
This is only because their more face-on counterparts hide behind populations of lower intrinsic ellipticity.

Another feature here is that massive galaxies can hardly reach the high ellipticity border of theoretical magenta lines, leaving some uncovered regions on the right.
This is probably the result of the growing significance of bulges in massive galaxies.
When viewed edge-on, their large bulges significantly reduce the ellipticity that you would measure out of their isophotes.
This makes them more inconsistent with the prediction of the magenta line, which assumes constant ellipticity.


\begin{figure*}

    \includegraphics[width=0.98\textwidth,trim={0 0.8cm 0 0},clip]{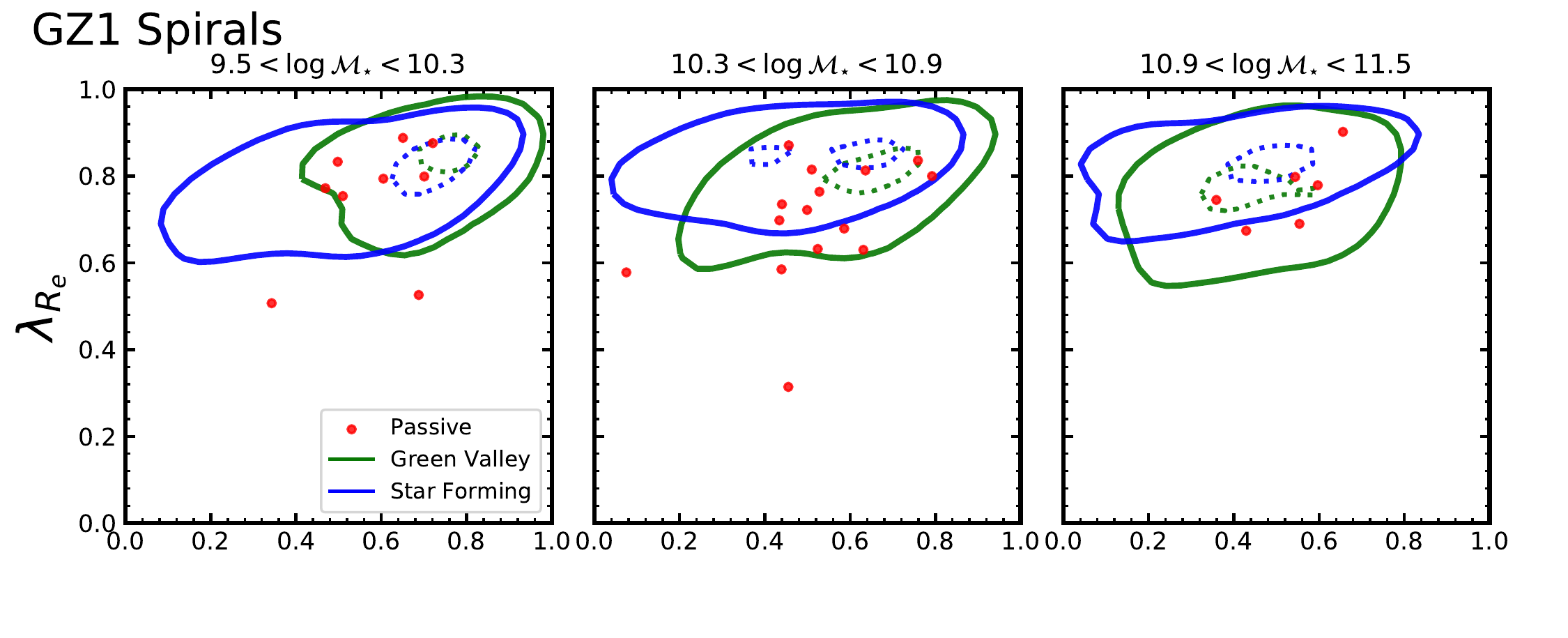}
  \includegraphics[width=0.98\textwidth,trim={0 0.8cm 0 0},clip]{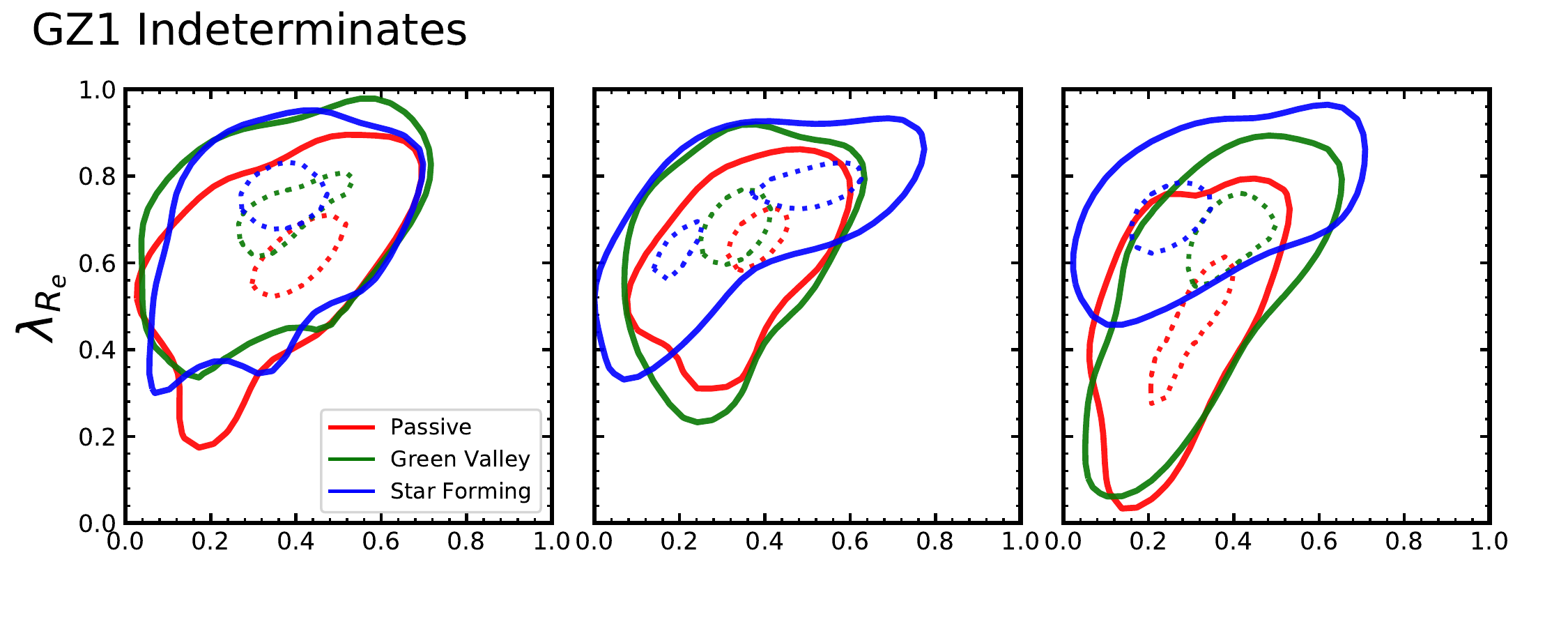}
  \includegraphics[width=0.98\textwidth]{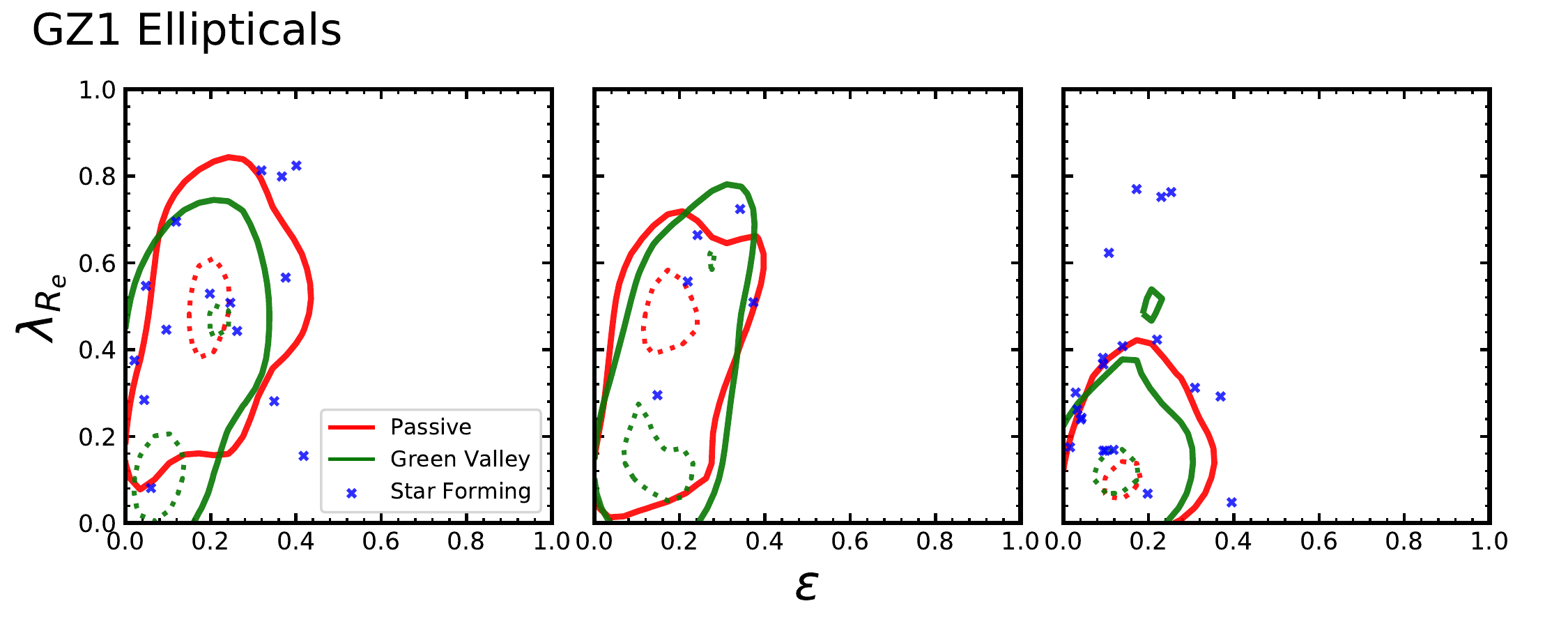}
    \caption{The same as in \autoref{fig:env2} but for Galaxy Zoo 1 classified spirals (top), indeterminate-type (mid) and elliptical galaxies (bottom).}
    \label{fig:morpho}
\end{figure*}


\subsection{Environmental dependence}\label{sec:envdep}

Apart from stellar mass, galaxy environment is another important driver of variation of galaxy properties.
In this section we investigate how the previous kinematic-state vs star-formation relation depends on environment indicated by a variety of environment diagnostics.

In left three columns of \autoref{fig:env1}, we illustrate $\lambda_{R_e}$ as a function of $\Delta _{\mathrm{MS}}$ for the same three mass slices, in which our data provide a large dynamic range of star formation rate.
From top down we show the dependence respectively on group identity (central/satellite), $\mathcal{N}$, $\Sigma_3$ and $\Sigma_{10}$ which are introduced in \autoref{subsec:env}.
The division according to group richness $\mathcal{N}$ is done by the median $\mathcal{N}$ of galaxies in groups ($\mathcal{N} \geq 2$).
While the dividing value 0.5 used for $\log\Sigma_3$  and  $\log\Sigma_{10}$ was chosen because it crudely corresponds on average to the separation between field ($\mathcal{N}=1$) and cluster ($\mathcal{N} \geq 20$) galaxies.
In each panel, the median $\lambda_{R_e}$ and the band enclosing 68\% of data points are shown.
In the last column, the median relations normalized to their values on SFMS are compared so that we get a sense how $\lambda_{R_e}$ varies with $\Delta _{\mathrm{MS}}$ in different stellar mass and environment bins.

A first result we can infer from this plot is the similarity between red and black shaded regions in all panels of left three columns.
This means at given stellar mass the decrease of $\lambda_{R_e}$ toward low $\Delta _{\mathrm{MS}}$ is not a strong function of environment.
By contrast, just as what has been shown in \autoref{fig:sfrm}, the relation varies substantially with mass.
From low to high mass, the difference between $\lambda_{R_e}$ on the SFMS and at low level of star formation increases apparently.
And in the highest mass bin, $\lambda_{R_e}$ sharply falls with a significant fraction of them condensed in a narrow region of low $\Delta _{\mathrm{MS}}$ and low $\lambda_{R_e}$ (note how close are the median and 16th percentile in the highest mass bin below $\Delta _{\mathrm{MS}}=-1$).

Although not significant, there is a hint of environmental dependence for the most massive galaxies.
As shown in the third column, below the SFMS central galaxies or the galaxies in denser regions or richer groups continuously have lower values of $\lambda_{R_e}$ over a large range of $\Delta_\mathrm{MS}$.
This means environment may indeed play a role in affecting kinematic properties of galaxies at least for massive galaxies.

Results described above is summarised in the last column of \autoref{fig:env1} where we see a monotonic change of relations with mass (from blue to red) while the split between galaxy populations in different environment (solid and dashed lines) is not comparably obvious.
From medium to high mass, the dramatic drop of $\lambda_{R_e}$ at a given $\Delta_\mathrm{MS}$ suggests the existence of mass threshold for low $\lambda_{R_e}$ systems.
Note that the reliability of group catalogue reduces at higher z.
So we also checked results by only looking at galaxies with $z<0.08$ and found results remained qualitatively the same.

\autoref{fig:kstest} confirms the signal of environmental dependence among massive galaxies by samples more rigorously matched in stellar mass and star formation level.
The first row shows the probability density distributions of $\mathcal{M}_{\star}$ (left) and $\Delta_\mathrm{MS}$ (middle) and $\lambda_{R_e}$ (right) for matched central and satellite galaxies, in the stellar mass range $10^{10.9}-10^{11.5} \mathcal{M}_{\odot}$ and with distance to the SFMS $\Delta_\mathrm{MS}<\,-0.35$ (where we observe the continuous $\lambda_{R_e}$ difference due to environment).
During matching procedure, for each satellite we randomly match it with a central galaxy with $\mathrm{log}\, \mathcal{M}_{\star}$ and $\Delta_\mathrm{MS}$ differing less than 0.02 and 0.1 respectively.
A more stringent mass control is given because the result that kinematics largely depends on mass.
Note that every central galaxy can only go into the final sample once.
And lastly we get matched central and satellite samples of 77 galaxies each.
We do the same for the other three environment indicators and get matched samples of 173 (group richness $\mathcal{N}$), 179 (local density $\Sigma_3$) and 137 ($\Sigma_{10}$) galaxies respectively, and the resultant distributions are shown in the following rows.

The left two columns show clearly that we have done well in controlling stellar mass and star formation level, indicated by difference of median values (marked by dashed lines) and the p-value of Kolmogorov-Smirnov (KS) test (shown at the upper right corner) which is the probability that the two distributions are drawn from the same underlying continuous distribution.
And last column confirms the environmental dependence of $\lambda_{R_e}$ for massive galaxies: the distribution of centrals or galaxies in larger groups or denser regions favours lower values of $\lambda_{R_e}$ and the p-values suggest statistical difference with confidence higher than 90\% for all except central/satellite dichotomy, for which we find a milder difference (72.5\% confidence).

As discussed previously, given large enough sample size along with the fact that the sample is not biased toward certain inclinations, comparing between average $\lambda_{R_e}$ alone is already informative.
Here by further taking ellipticity into account, we show results in \autoref{fig:env2} that are consistent with our aforementioned findings.
For brevity, among those environment indicators we only show the results with central/satellite dichotomy (upper and lower row respectively) and we confirmed that the conclusion remain the same for other environmental indicators.
Again, galaxies are split into three mass bins and in each bin we plot on $\lambda_{R_e} - \epsilon$ plane the KDE derived probability density distribution of star-forming, green valley and passive galaxies separately.
The contour enclosing 68 per cent of total probability is denoted as a solid line illustrating the position of bulk distribution, and the one enclosing 10 percent as a dotted line marking the position of peak.

For both rows, the distributions of star-forming galaxies all look almost identical.
When stellar mass increases, the distributions of the green valley and in particular the passive galaxies, migrate to the region of small $\lambda_{R_e}$ and $\epsilon$.
In the highest mass bin, for green valley and passive galaxies, the distributions of central galaxies become more concentrated in the small $\lambda_{R_e}$ and $\epsilon$ region than the satellites, indicating an environmental dependence.
While there is no obvious trend seen for the two bins of lower mass.
All these results are consistent with those in \autoref{fig:env1}.


\begin{figure*}
    \includegraphics[width=0.95\textwidth]{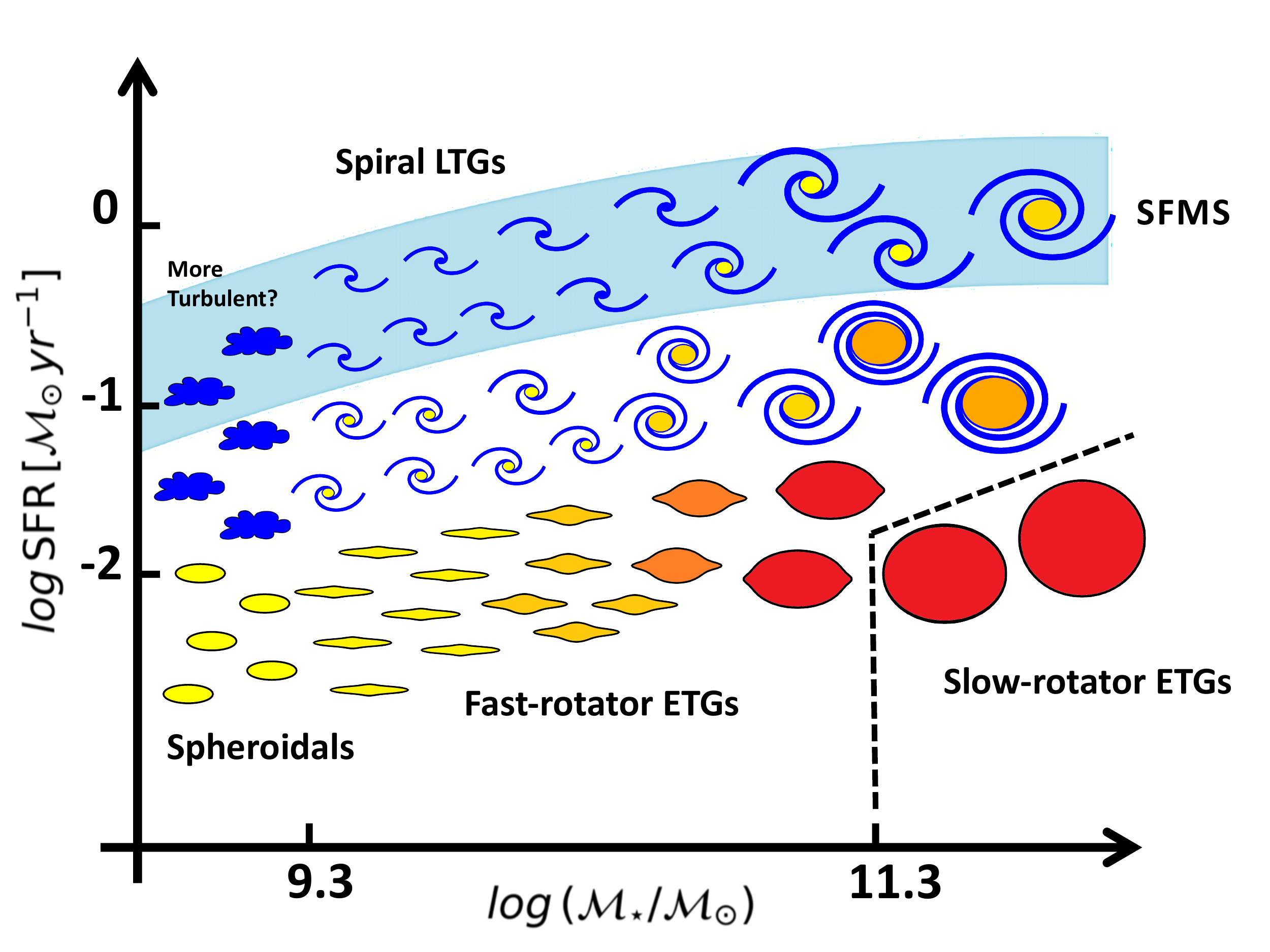}
    \caption{Schematic plot for \autoref{fig:sfrm}.
        The blue band denotes the span of the SFMS ($\pm 0.35 $ dex).
        Above $10^{9.3} \mathcal{M}_{\odot}$ galaxies with star formation are illustrated in a relatively face-on view and with spiral arms and different size of bulges (denoted ``Spiral LTGs'').
        While star-forming galaxies at the very low mass end are shown by symbols with more irregular morphology which is hinted at by their lower values of $\lambda_{R_e}$.
        For passive galaxies, we show them as they are seen edge-on in order for a clearer view of disk significance.
        We emphasize the mass threshold of $10^{11.3} \mathcal{M}_{\odot}$ beyond which the genuine slowly rotating ETGs (``Slow-rotator ETGs'') are preferentially found.
        And other passive galaxies with lower mass are denoted ``Fast-rotator ETGs'' and ``Spheroidals'' below $10^{9.3} \mathcal{M}_{\odot}$. The symbols in this figure were chosen to be the same as in the related fig.~23 and 24 of \citet{2016ARA&A..54..597C}, where one can find further details on the definitions.
         }
    \label{fig:scheme}
\end{figure*}


\subsection{Link with visual morphology classifications}

Here we check how visually classified morphology compares to the view from kinematics by taking a quick look at the distributions of GZ1-S, GZ1-I and GZ1-E galaxies on $\lambda_{R_e} - \epsilon$ plane.
Among $\mathcal{S}_{\mathrm{MaNGA}}$ galaxies with mass falling into the range $10^{9.5}-10^{11.5} \mathcal{M}_{\odot}$, 37.6\%, 39\% and 23.4\% are GZ1-S, GZ1-I and GZ1-E respectively.

The results are shown in \autoref{fig:morpho} in the same manner as in \autoref{fig:env2} with galaxies split into GZ1-S (top row), GZ1-I (mid row) and GZ1-E (bottom row).
As there are only a handful of passive GZ1-S and star-forming GZ1-E in $\mathcal{S}_{\mathrm{MaNGA}}$, instead of showing their KDE derived density distribution we directly show each individual point on the plane.

First, we note that the distributions of GZ1-S and GZ1-E have little overlap which indicates that they are dynamically distinct populations.
If LTGs evolve and become ETGs directly, then this can mean the change of their kinematics must be very fast.
Alternatively, LTGs can first gradually evolve into a transitional phase, and then become ETGs later without rapid change of their kinematics.
Therefore, it is interesting to notice that the GZ1-I right occupy the transition regions between GZ1-S and GZ1-E on the $\lambda_{R_e} - \epsilon$ plane.
Though the physical nature of these GZ1-I is still unclear (partially due to the poor image quality of SDSS) they may indeed represent a transitional population between GZ1-S and GZ1-E.

On the other hand, we stress that within certain morphological population and mass bin the distributions of galaxies with different star formation levels look very similar in general.
This implies that star formation quenching processes are not necessarily accompanied by the change of morphology and kinematics.
For instance, as the top row suggests, the star-forming LTGs can be quenched and become passive LTGs with similar level of rotation.
We will further explore the interrelationship between star formation, morphology and kinematics in our future work.

The position of the distributions of GZ1-S, GZ1-I and GZ1-E on the $\lambda_{R_e} - \epsilon$ plane indeed indicates a strong connection between galaxy morphology and kinematics.
But it is clear that relatively face-on passive disks are missed by visual classifications GZ1-S as opposed to their edge-on counterparts.
As the high $\lambda_{R_e}$ and low $\epsilon$ region is barely populated by passive GZ1-S in top row where we expect those relatively face-on disks to be (see also the analytic tracks of changing inclination in \autoref{fig:le}).
And part of these face-on disks have probably been classified as GZ1-E which can be found in the high $\lambda_{R_e}$ and low $\epsilon$ region in the bottom row.
This echos the result of $\mathrm{ATLAS}^{\mathrm{3D}}$ which shows that the vast majority of early-type galaxies are actually fast rotating \citep{2011MNRAS.414..888E,2011MNRAS.414.2923K}.
And it also shows the strong dependence of visual classifications on inclination.

\section{Discussion}

Probably the most striking result of this paper is \autoref{fig:sfrm}.
It shows the dramatic mass trend of the kinematic T-C relation and emphasizes that relatively low-mass ($\sim 10^{9.8} \mathcal{M}_{\odot}$) passive galaxies are generally disk dominated systems.
This is against the direct impression from their images because when inclined and passive, in many cases they look roughly the same as spheroids (especially when the resolution of images is low).
Indeed there have been evidence showing the mass dependence of rotation in elliptical galaxies as already mentioned in the introduction section, in this figure we take a step forward by connecting together mass, star formation level and kinematics.

In \autoref{fig:scheme} we present a cartoon version of \autoref{fig:sfrm} illustrating the main results and making a direct link between angular momentum $\lambda_{R_e}$ and galaxy morphology.
Galaxies are put on $\mathrm{SFR}-\mathcal{M}_{\star}$ diagram where we use a blue band to show the span of the SFMS.
To distinguish between passive galaxies and those with star formation and also to highlight the disk nature of relatively low-mass passive galaxies, we use different types of symbols. We chose the symbols to be the same as fig.~23 and 24 in the review by \citet{2016ARA&A..54..597C}. This makes it easier for the reader to understand the connection between galaxy properties on the $\mathrm{SFR}-\mathcal{M}_{\star}$ and in other related diagrams.

For passive galaxies, we illustrate them as they are viewed edge-on to make their structure recognizable.
Except those at the very low mass end, galaxies with star formation are shown as systems with spiral arms with different size of bulges.
While we put symbols with more irregular morphology at the low mass end to illustrate the lower $\lambda_{R_e}$ there and it is indeed supported by visual check of their SDSS images many of which reveal irregular features.
Stronger turbulence may be the reason for their lower values of $\lambda_{R_e}$ and it will be discussed in the following section.
Even though our data include almost no passive galaxies at this very low mass end (possibly due to the MaNGA magnitude-based selection), for completeness, we still put symbols for spheroidal galaxies in this part of parameter space, based on previous results.
As they are dominating the passive population at this low mass in the local universe and there is convincing empirical evidence showing their connection with dwarf irregular galaxies \citep{2012ApJS..198....2K}.
Given the weak star formation dependence of $\lambda_{R_e}$ among low-mass galaxies, the evolutionary link between spheroidals and not-so-cold low-mass star-forming systems seems natural.

Note that here, for the slight decrease of $\lambda_{R_e}$ along the SFMS at the high mass end, we represent this as due to an increase of the bulges, following the known link between $\lambda_{R_e}$ and bulge fraction \citep[e.g., sec.~3.6.3 of][]{2016ARA&A..54..597C}.
The fact that we observe a slight decrease of $\lambda_{R_e}$ only at the high mass of the SFMS is consistent with previous observations of a trend of increasing bulge to total ratio with mass in that region of the diagram \citep[e.g.,][]{2019MNRAS.485..666B}.

To conclude, this figure highlights the similarity between low-mass star-forming and passive galaxies in terms of the significance of disk components.
While it is opposed to the progressively more dominating spheroid components in massive passive galaxies and beyond $10^{11.3} \mathcal{M}_{\odot}$, the disk-free genuinely spheroidal slowly rotating ETGs are preferentially found.

The spin parameter used throughout this work is measured within the effective (half-light) ellipse and it may vary using a larger aperture. The half-light ellipse is used because any classification needs to be define within a physical scale, and the $\lambda_{R}$ measured within that radius was shown to be easy to measure for large samples and to correlate well with other galaxy properties \citep[see review by][]{2016ARA&A..54..597C}.
However, given the relatively flat $\lambda_{R}$ profiles beyond $R_e$ (See Fig.5 of \citealt{2014ApJ...786...23R}, Fig.5 of \citealt{2016MNRAS.457..147F} and Fig.14 of \citealt{2017MNRAS.471.4005B}), the kinematic classification in most cases does not changes when using different radii and our conclusions would have remained qualitatively the same if we had adopted a less-optimal and larger aperture.

The same dependence on the chosen aperture applies to our determination of the SFR. For this, in \aref{app:repro}, we have shown that the result stays qualitatively the same after we replace the total SFR shown in \autoref{fig:sfrm} by a SFR measured in a completely different manner, using the H$\alpha$ fluxes derived directly from the MaNGA data, within one $R_e$.

Galaxy populations produced by the state-of-the-art cosmological hydrodynamic simulations have shown qualitatively similar behaviour in terms of the lower level of rotation for star-forming galaxies both at lower and higher mass end, and also the fact that lowest spin is owned by the most massive passive galaxies \citep{2017MNRAS.472L..45C,2018MNRAS.476.4327L}.
But we note that simulated passive galaxies with $\mathcal{M}_{\star} \sim 10^{10} \mathcal{M}_{\odot}$ are more dispersion dominated than the observed ones.

It is revealing to compare our \autoref{fig:sfrm} and \autoref{fig:scheme} with the left panel of Fig.~1 of \citet{2011ApJ...742...96W}, which visualizes the distribution of the \citet{Sersic1968} index on the $\mathrm{SFR}-\mathcal{M}_{\star}$ diagram. That diagram is often presented to demonstrate the connection between galaxy structure and SFR, with the passive galaxies having a index close to the \citet{deVaucouleurs1948} profile ($n=4$) while the star forming ones having exponential profiles ($n=1$) typical for disks. This is interpreted as galaxies generally becoming ellipticals when they quench and become passive.

Our diagrams show a similar feature. However there is a crucial difference. In fact, while in the \citet{2011ApJ...742...96W} figure the region of ``ellipticals'' span the entire mass range from just above $\mathcal{M}_\star\sim10^{10}$ $\mathcal{M}_\odot$ up to the largest masses of nearly $\mathcal{M}_\star\sim10^{12}$ $\mathcal{M}_\odot$, in our diagrams there is a clear transition at a significantly larger mass of $\mathcal{M}_\star\sim10^{11}$ $\mathcal{M}_\odot$. Only above that mass being passive is associated to being slow-rotator ``ellipticals'', while for lower masses quenching produces no more than a minor structural transformation (increasing bulges but retaining significant level of rotation). This is the key novelty we want to emphasize with this work.

A transition around $\mathcal{M}_\star\sim10^{11}$ $\mathcal{M}_\odot$ is not unexpected, given that only above that critical mass are slow-rotator ETGs starting to dominate, while below that mass one only finds spiral galaxies and fast rotator ETGs \citep[e.g.][]{2016ARA&A..54..597C}. However, in this paper we first illustrate this fact specifically in the context of the $\mathrm{SFR}-\mathcal{M}_{\star}$ diagram, where this fact is not yet universally appreciated.

Beyond fitting the light distribution with single Sersic profiles, efforts have been made to decompose galaxies into multiple components including bulges and disks.
A relevant result in \citet{2017A&A...597A..97M} illustrating bulge to total ratio as a function of stellar mass and SFR shows overall consistency with what we found using the spin parameter.
The same difference appears in the passive population for which the spin reveals a sharp transition among massive galaxies whereas Fig. 5 of \citet{2017A&A...597A..97M} by contrast indicates very close bulge to total ratio over a large range of mass.

In the following section, we discuss the implications of our results in the context of a $\Lambda \mathrm{CDM}$ universe in order to understand the trend of $\lambda_{R_e}$ with stellar mass and star formation level.

\subsection{Understanding in the context of galaxy formation and evolution within a $\Lambda \mathrm{CDM}$ universe}

\subsubsection{Early/late star formation and in/ex situ mass assembly}

To understand the current kinematics of galaxies, we need to know what factors are key to determining their kinematic state.
In principle, whether a galaxy is mainly in ordered rotation or random motion (that is disk-dominated or spheroid-dominated) will depend on 1) the angular momentum of gas used as raw material for making stars and 2) how these stars are assembled.
Just as shown in \citet{2019ApJ...874...67B} for a Milky Way like disk galaxy in a cosmological hydrodynamical simulation, stars with lower initial birth angular momentum end up in a non-rotating spherical bulge.
But even stars are born with high angular momentum, if they are assembled via violent mergers, the final aggregate is still likely to be dominated by random motion.

These two aspects are translated into two crucial factors of galaxy formation: 1) formation time of stars of the galaxy and 2) the fractional stellar mass assembled in situ and ex situ.
Note that the formation time of stars of the final galaxy is not necessarily close to the assembly time of this galaxy.
As stars can form in separate galaxies at higher redshift while they merger at low redshift to make up the final galaxy \citep{2010ApJ...725.2312O}.

The origin of angular momentum of galaxies remains a hot topic in the field.
To first order, it is believed that protohalos (including their gas) gain angular momentum through tidal torques from their environment until maximum expansion (turnaround), and subsequently detach with Hubble flow and collapse into virialized structures that preserve their angular momentum \citep{1969ApJ...155..393P, 1970Afz.....6..581D, 1984ApJ...286...38W, 1996MNRAS.282..436C, 1996MNRAS.282..455C}.
Before turnaround, angular momentum in protohalos grows linearly with time.
Therefore, stars formed earlier than turnaround can have lower specific angular momentum than those formed later, contributing to spheroidal component \citep{2016MNRAS.460.4466Z,10.1093,2020arXiv200310912R}.
On the other hand mergers, i.e. ex situ mass assembly, tend to randomise the stellar kinematics via violent relaxation and result in a transfer of angular momentum from galaxies to outer halos \citep{1977egsp.conf..401T, 1992ApJ...400..460H, 2006ApJ...636L..81N}.
Thus, the angular momentum gained via tidal torques during the linear growth stage can only be preserved when stars are assembled in situ, i.e. star formation on the cold gas disks \citep{1980MNRAS.193..189F, 1998MNRAS.295..319M}.
This explains the importance of the second factor put forward in last paragraph.
One thing that has not been included explicitly in the second factor is the wetness of mergers.
A significant amount of gas involved in dissipative mergers can suppress growth of boxy orbits \citep{2006MNRAS.372..839N} and can also re-build a disk \citep{2009ApJ...691.1168H}.
But given that the subsequent stellar disk builds out of the survived gas disk by in situ star formation, the second factor has already implicitly taken this into account.
This means when we say a galaxy has a large fraction of stars assembled ex situ, it directly rules out the case where much of its stellar mass is formed out of gas brought by a wet merger and it indicates that dry mergers dominate.

As a result, a galaxy is more likely to be disk-dominated when most of its stars are assembled in situ at lower redshift
\footnote{But rarer cases do happen where a minor merger completely destroys the pre-existing disk \citep{2019MNRAS.489.4679J} or where a significant fraction of accreted gas has largely misaligned angular momentum \citep{2012MNRAS.423.1544S}.
And in these cases though little mass is assembled ex situ, galaxies can end up with spheroids.}
which echos what is concluded for disk formation in \citet{2017ARA&A..55...59N} that to form a disky galaxy in simulation early star formation should be suppressed by effective feedback so that ejected gas can fall back to the galaxy with higher angular momentum \citep{2007MNRAS.374.1479G, 2014MNRAS.443.2092U}.
While a galaxy that either assembles much mass via accreting existing stars or converts gas into stars at early universe tends to end up being spheroid-dominated.
What outlined here is much in line with the conclusion in \citet{2017MNRAS.464.3850L}.
Using state-of-the-art cosmological hydrodynamical simulation, they found galaxy mergers and early star formation quenching (thus most of stars form at high redshift) as two primary channels to galaxies with low specific angular momentum.

\subsubsection{The mass dependence of kinematic-morphology vs star-formation relation}

With these two factors in mind, now we attempt to understand the result that among massive galaxies star-forming ones are rotating whereas passive ones are not and low mass galaxies are mainly disk-dominated regardless of star formation.

In the currently accepted paradigm of the $\Lambda \mathrm{CDM}$ universe, gas cools and condenses into stars at the bottom of dark matter halos which are the result of gravitationally amplified primordial density perturbations.
Highest density peaks grow rapidly toward massive halos and they later form the central part of clusters after hierarchical merging.
Within these most massive halos, the evolution of the galaxies proceeds clearly in two phases \citep{2007ApJ...658..710N, 2010ApJ...709..218F, 2010ApJ...725.2312O, 2012ApJ...754..115J, 2017MNRAS.464.1659Q}.
At high redshifts ($z\, \gtrsim\, 2$), galaxies grow by in situ star formation.
The star formation is so intense that these galaxies assemble a large amount of stellar mass on short timescales of the order of one Gyr \citep{2002ApJ...576..135G, 2004Natur.428..625H, 2005ApJ...621..673T, 2010MNRAS.404.1775T, 2015MNRAS.448.3484M}.
And it is out of gas with relatively low angular momentum such that some galaxies can be extremely small and compact \citep{2010ApJ...722.1666W, 2011ApJ...730....4B, 2010ApJ...725.2312O, 2012ApJ...744...63O, 2016MNRAS.456.1030W}.
Later in situ star formation is suppressed as cold gas is shock heated to high temperature  and thus it can no longer easily penetrate through hot halos \citep{2005MNRAS.363....2K, 2006MNRAS.368....2D}.
And subsequently AGN radio mode feedback becomes effective \citep{2013ARA&A..51..511K}.
At this stage mass assembly is switched to ex situ mode and galaxies further grow by accreting stars formed in other galaxies (mergers tend to be dry due to massive halos).
And it has been reported that accreted mass on average makes up a half of these massive galaxies \citep{2014ARA&A..52..291C}.
This two-phase characteristic is particularly true for passive galaxies because they reside in more massive halos than their star-forming counterparts, which has been shown by weak gravitational lensing \citep{2006MNRAS.372..758M, 2016MNRAS.457.3200M}.
Therefore, it can be understood why present-day massive passive galaxies have their spheroid-dominated kinematics according to the two factors discussed previously.

For galaxies with lower mass, cosmological simulations clearly indicate that stellar accretion is less important \citep{2010ApJ...725.2312O, 2012MNRAS.425..641L} so that in situ star formation prevails.
On the other hand, low-mass systems form later \citep[i.e. with younger age; ][]{2005MNRAS.362...41G}.
This so-called downsizing \citep{1996AJ....112..839C, 2006MNRAS.372..933N} is partly due to more effective feedback that delays star formation in shallower potential wells \citep{2002MNRAS.335..487M}.
Together, ubiquitous disks shown among low-mass MaNGA galaxies are expected because of both more disk construction (later gas infall) and less disk destruction (fewer mergers).
However, exception exists at the very low mass end of the SFMS ($\mathcal{M}_{\star} \sim 10^{9.2} \mathcal{M}_{\odot}$).
Just like their more massive star-forming counterparts, these low-mass star-forming galaxies assembled most of mass at low redshifts \citep{2007MNRAS.381..263A, 2010ApJ...721..193P, 2012ApJ...745..149L} via gas accretion.
But they do not show significant rotation as more massive star-forming galaxies do, which is likely due to their large cold gas reservoir.
For these systems, the mass of cold atomic gas is already about three times the mass of stars \citep{2012ApJ...756..113H}.
And the strong specific gas accretion rate responsible for this into a shallow potential well may disturb disks and boosts turbulence, reminiscent of high-redshift turbulent galaxies \citep{2008ApJ...687...59G, 2012Natur.487..338L}.
Supernova feedback can also play an important role in disrupting existing gas disks as well as preventing the supply of high angular momentum gas for disk construction \citep{2020MNRAS.493.4126D}.
This suggests a different assembly mode among these less massive systems \citep{2018MNRAS.478.3994C}.

\subsubsection{Other factors}

Above arguments qualitatively explains the relation of kinematics and star formation we observe as a function of stellar mass.
Nevertheless, kinematic and morphological evolution of galaxies involve much more than what is discussed above.

There are other important internal processes that also play a role.
Large Jeans mass on high-redshift turbulent disks causes fragmentation into massive clumps which migrate toward central region under dynamical friction and contribute to classical bulge growth \citep{2014MNRAS.438.1870D, 2015MNRAS.450.2327Z}.
This may be also partly responsible for kinematics of present-day massive passive galaxies.

\citet{2004ARA&A..42..603K} argues while hierarchically driven galaxy evolution is more dominant in the past, now secular internal evolution is gradually taking over.
Among large non-axisymmetric structures, bars are known to enhance galaxy mass concentration and buckle to make disks thicker \citep{2004ARA&A..42..603K}.
But these instabilities usually serve as mild negative feedback to adjust galaxies rather than altering them completely.
For example, bars grow in systems that are dynamically too cold and with mass concentration too low.
They fuel central star formation by funnelling gas in and heat the disk vertically under buckling instability but increased mass concentration in turn dissolves the bar \citep{1990ApJ...361...69H, 1993A&A...268...65F, 1999ApJ...510..125S, 2004ApJ...604..614S}, which renders the effects of bar less overwhelming.

Externally, it is also reported that tidal interaction with neighbours in group environment can modify the shape of galaxies \citep{1979MNRAS.188..273B, 2006PASP..118..517B}.
However, except the most massive galaxies we do not see apparent dependence of $\lambda_{R_e}$ on environment at a given star formation level.
Though we cannot exclude the possibility that these tidal interactions move galaxies along the median relation, i.e. reducing $\lambda_{R_e}$ and SFR together along the track of central galaxies.
For the most massive galaxies, we do identify a trend that random motion dominates more if galaxies are central/in denser environment/in richer groups, at the same star formation level.
This is qualitatively consistent with merging is strongest for central galaxies in galaxy clusters \citep{1977ApJ...217L.125O, 2007MNRAS.375....2D}.
And it confirms that at given stellar mass, kinematic state indeed depends on environment as opposed to the purely mass-driven kinematic trend in some studies based on smaller dataset \citep[e.g.,][]{2017ApJ...851L..33G, 2017ApJ...844...59B}.

One may also resort to the properties of dark matter halos to understand galaxies considering the close connection between the halo and the baryonic component \citep{2018ARA&A..56..435W}.
Assuming that dark matter halo and gas share the same specific angular momentum, galaxy populations with realistic properties can be reproduced with the knowledge of spin distribution from numerical simulations \citep{1997ApJ...482..659D,1997MNRAS.292L...5J,1998MNRAS.295..319M,1998ApJ...507..601V}.
And those low spin halos may partly responsible for the formation of bulges and spheroidal galaxies.
But clearly there is gap between the numbers of dispersion supported galaxies and halos at the low spin tail of distribution \citep{2012ApJS..203...17R}.
And the spin distribution of halos from simulations does not depend on the mass of halos \citep[e.g.,][]{2002MNRAS.329..423M}\footnote{Despite of some observational evidence against this \citep{2008MNRAS.388..863C,2008MNRAS.391..197B}.}, in contrast with the dramatic mass trend that we found for galactic spin parameter.
These suggest more roles played by baryonic physics, such as the heating from stellar feedback that prevents over-cooling and its subsequent massive transfer of angular momentum from baryons to halo particles \citep{2002MNRAS.335..487M}, which can lead to a mass dependence of stellar angular momentum.

\subsection{Implications for quenching mechanisms}

From our results we may also be able to draw some implications for how star formation is getting or has got quenched.

For massive galaxies, as discussed above the decrease of SFR together with the shrinking significance of stellar disk can be linked by progressively more massive dark matter halos.
Presence of a massive halo implies a merger-rich assembly history as well as suggesting the absence of cold streams.
And an extremely super massive black hole as the outcome of abundant mergers keeps the circumgalactic medium hot.

By contrast, the disk dominance present even among the reddest low-mass galaxies implies that those contributing factors to the quenching of low-mass galaxies do not severely affect their kinematics.
Without doubt, one would immediately think of environment-related mechanisms which are believed to be effective among low-mass galaxies.
For instance, various kinds of hydrodynamical interaction with hot intergalactic medium such as ram pressure stripping \citep[e.g.,][]{1972ApJ...176....1G, 1999MNRAS.308..947A, 2000Sci...288.1617Q} and strangulation \citep[e.g.,][]{1980ApJ...237..692L, 2015Natur.521..192P} can largely affect gas disks while leaving stellar disks almost intact.
However, considering that these mechanisms are mainly efficient in massive halos there comes a question that why median $\lambda_{R_e}$ at a given star formation level does not differ for low-mass galaxies in different environment.
A straightforward solution would be that the quenching mechanisms responsible for star formation cessation in relatively isolated galaxies also do little on the kinematics.
Thus commonly proposed reionization and stellar feedback \citep{1986ApJ...303...39D, 2000ApJ...539..517B, 2002ApJ...572L..23S} as primary quenching mechanisms for low-mass galaxies are indeed feasible from this stand point.

\subsection{Comparing with relevant IFU studies}

So far the extraction of conclusion for many of our results has relied on the statistical significance given by MaNGA data release 15.
For example, such large sample allows for a fairly good coverage on $\mathrm{SFR}-\mathcal{M}_{\star}$ plane down to a relatively low mass which has helped to reveal the striking mass trend of kinematics.
Additionally, it also sheds light on those straggling counter-rotating populations on $\lambda - \epsilon$ plane that would otherwise be missed by smaller samples due to the rarity.
In this section, we briefly compare our \autoref{fig:le} and \autoref{fig:env1} with the main results in \citet{2018NatAs...2..483V} and \citet{2019MNRAS.485.2656C} respectively given they are highly relevant.

\citet{2018NatAs...2..483V} illustrates a good correlation between intrinsic ellipticity (as inferred on $V/\sigma - \epsilon$ diagram) and luminosity-weighted age of $\sim\, 800$ galaxies from Sydney–Australian Astronomical Observatory Multi-object Integral field spectrograph (SAMI) Galaxy Survey \citep{2012MNRAS.421..872C}, which is similar to what we observe for intrinsic ellipticity and star formation level.
The consistency is marked especially considering that how much the difference is between their stellar age and the SFR that we use.
They take the age of a representative single stellar population model that best reproduces the measured Lick indices within aperture of radius one $R_e$ \citep{2017MNRAS.472.2833S}, while SFR here is the one averaged over the past 100-Myr star formation history of the best fit stellar population synthesis model, fitted to total light (rather than in an aperture).
The "stragglers" we found on $\lambda - \epsilon$ plane barely have counterparts on $V/\sigma - \epsilon$ plane in \citet{2018NatAs...2..483V}, which is very likely due to their relatively small sample size.

The main result of \citet{2019MNRAS.485.2656C} (their Fig. 3) shows the variation in stellar $V/\sigma$ as a function of star formation level for $\sim 170$ satellite galaxies with SAMI IFS data, resembling what we have done for MaNGA satellite galaxies in the first row of \autoref{fig:env1}.
Because in their result the average change is not far from zero for a large range of star formation level, they conclude that satellite galaxies undergo little structure change during quenching phase.
However, from our results the structure change as indicated by $\lambda_{R_e}$ does not highly depend on environment but stellar mass.
And indeed when looking at Fig. 3 of \citet{2019MNRAS.485.2656C}, there is detectable mass trend at a given star formation level in the same sense as we found in our data.
So their mean relation being close to zero is due to the fact that the result is relatively dominated by less massive galaxies ($\mathcal{M}_{\star} \lesssim 10^{10.5} \mathcal{M}_{\odot}$).
It is true that we have analyzed in a slightly different way and particularly they match ellipticity when getting $\Delta _{V/\sigma}$.
But as we have shown that our conclusion does not change when further taking ellipticity into account (\autoref{fig:env2}), this difference should not be a key point.

\section{Summary}

In this work, using the spin parameter $\lambda_{R_e}$ as a kinematic indicator of disk significance we have studied the correlation between morphology and star formation level and its dependence on stellar mass and environment for $\sim 3200$ MaNGA galaxies.
Such sample size together with the (nearly) mass-based selection scheme of MaNGA survey allow for a good coverage on $\mathrm{SFR}-\mathcal{M}_{\star}$ plane down to about $10^{9.5}$ solar mass.
And our findings are summarized as follows:

\begin{enumerate}[(i)]
  \item The SFMS is approximately a ridge of maximum values for the specific angular momentum $\lambda_{R_e}$ distribution on the $\mathrm{SFR}-\mathcal{M}_{\star}$ plane. In fact, $\lambda_{R_e}$ decreases toward both low and high SFR, at given stellar mass.
  On the SFMS galaxies have large specific angular momentum ($\lambda_{R_e}\sim 0.7$), or equivalently are dominated by rotation, and there is little change in $\lambda_{R_e}$ with mass, with the exception of the lowest masses ($\mathcal{M}_{\star} \lesssim 10^{9.3} \mathcal{M}_{\odot}$), where $\lambda_{R_e}$ decreases.

  \item Below the SFMS, and for $\mathcal{M}_{\star}\gtrsim10^{11} \mathcal{M}_{\odot}$, there is a dramatic change in $\lambda_{R_e}$ with SFR ($\Delta\,\lambda_{R_e}\sim 0.5$) and passive galaxies are dominated by slow-rotator ETGs. When $\mathcal{M}_{\star}\lesssim10^{11} \mathcal{M}_{\odot}$ the variation of $\lambda_{R_e}$ with SFR becomes more marginal and passive galaxies are mainly fast-rotator ETGs, which have a significant amount of rotation.
  And particularly at low masses ($\mathcal{M}_{\star} \sim 10^{9.8} \mathcal{M}_{\odot}$) on average $\Delta\,\lambda_{R_e}$ has a value of only about 0.1.

  \item Such progressively tighter relation between $\lambda_{R_e}$ and $\mathrm{SFR}$ toward high stellar mass displays as clearer stratification pattern on the $(\lambda_{R_e},\epsilon)$ plane, in the sense that galaxy population with a certain star formation level matches the theoretical track, for different inclinations, of galaxy population of certain intrinsic ellipticity.
  This means for massive galaxies intrinsic morphology is a good indicator of star formation state.

  \item Counterrotating stellar disks are clear outliers on the $(\lambda_{R_e},\epsilon)$ plane in the lower right region (relatively high apparent ellipticity but low $\lambda_{R_e}$).
  They stand out for their star formation level larger than what one would expect given their $\lambda_{R_e}$. This is because their low $\lambda_{R_e}$ is due to counterrotating stellar disks and not to a spheroidal non-rotating morphology.

  \item The environmental dependence of $\lambda_{R_e}-\mathrm{SFR}$ relation is much weaker than its mass dependence but it is unambiguous for the massive galaxies.
  In our highest mass bin $10^{10.9}-10^{11.5} \mathcal{M}_{\odot}$, central galaxies or the galaxies in dense regions or rich groups on average have a lower value of $\lambda_{R_e}$. While no obvious environmental dependence is found for low-mass galaxies.

  \item The position of the distributions of spiral, elliptical and indeterminate-type galaxies (according to Galaxy Zoo 1 visual classifications) on $(\lambda_{R_e} , \epsilon)$ plane suggests a strong connection between galaxy morphology and kinematics.
  In particular, the indeterminate-type galaxies right occupy the region between spiral and elliptical galaxies and hence may represent an transitional population between the two.
  Together they form a continuous sequence which may also imply an evolutionary track of galaxies in their morphology and kinematics.
  However it is clear that a fraction of relatively face-on passive disks are visually  classified into a different category compared with their edge-on counterparts.
  This shows the strong dependence of visual morphological classifications on inclination and reflects the advantage of using IFS data to probe galactic structure.

\end{enumerate}

These results have shown that a good correspondence between star formation level and intrinsic morphology is only seen for massive galaxies ($\mathcal{M}_{\star} \gtrsim 10^{10.5} \mathcal{M}_{\odot}$).
And the traditional picture that galaxies below the SFMS are generally spheroidal or bulge-dominated has been shown not to apply for galaxies with lower mass ($10^{9.5} \lesssim \mathcal{M}_{\star} \lesssim 10^{10} \mathcal{M}_{\odot}$).
As among them, even the reddest (with oldest stellar populations and lowest SFR) are still disk-dominated fast-rotator ETGs.
This concurs with the fact found by $\mathrm{ATLAS}^{\mathrm{3D}}$ survey that two thirds of visually classified elliptical galaxies (which are thought to be spheroidal or ellipsoidal in 3D space) are fast rotators thus with significant disk component.

We have discussed that this mass and star formation dependence of disk significance can be qualitatively understood in the context of formation time of stars and mass assembly mode.
A galaxy that either assembles much mass via accreting existing stars or has most of its stars formed at early universe tends to end up being spheroid-dominated.
While a galaxy is more disk-dominated if it assembles mainly via in situ star formation fuelled by gas accretion at later time.
And the flat $\lambda_{R_e}-\mathrm{SFR}$ relation of less massive galaxies together with its negligible dependence on environment suggest that though the quenching mechanisms are likely to be different for group and field low-mass galaxies, they all have little effect on galaxy kinematics.

\section*{Acknowledgements}

BW is grateful for the financial support from China Scholarship Council during his stay in Oxford.
YP acknowledges the National Key R\&D Program of China, Grant 2016YFA0400702 and NSFC Grant No. 11773001, 11721303, 11991052.

Funding for the Sloan Digital Sky Survey IV has been provided by the Alfred P. Sloan Foundation, the U.S. Department of Energy Office of Science, and the Participating Institutions. SDSS acknowledges support and resources from the Center for High- Performance Computing at the University of Utah. The SDSS website is \url{www.sdss.org}.

SDSS-IV is managed by the Astrophysical Research Consortium for the
Participating Institutions of the SDSS Collaboration including the
Brazilian Participation Group, the Carnegie Institution for Science,
Carnegie Mellon University, the Chilean Participation Group, the French Participation Group, Harvard-Smithsonian Center for Astrophysics,
Instituto de Astrof\'isica de Canarias, The Johns Hopkins University, Kavli Institute for the Physics and Mathematics of the Universe (IPMU) /
University of Tokyo, the Korean Participation Group, Lawrence Berkeley National Laboratory,
Leibniz Institut f\"ur Astrophysik Potsdam (AIP),
Max-Planck-Institut f\"ur Astronomie (MPIA Heidelberg),
Max-Planck-Institut f\"ur Astrophysik (MPA Garching),
Max-Planck-Institut f\"ur Extraterrestrische Physik (MPE),
National Astronomical Observatories of China, New Mexico State University,
New York University, University of Notre Dame,
Observat\'ario Nacional / MCTI, The Ohio State University,
Pennsylvania State University, Shanghai Astronomical Observatory,
United Kingdom Participation Group,
Universidad Nacional Aut\'onoma de M\'exico, University of Arizona,
University of Colorado Boulder, University of Oxford, University of Portsmouth,
University of Utah, University of Virginia, University of Washington, University of Wisconsin,
Vanderbilt University, and Yale University.




\bibliographystyle{mnras}



\appendix

\section{Star Formation Main Sequence}\label{app:sfms}

\begin{figure}
    \includegraphics[width=0.45\textwidth]{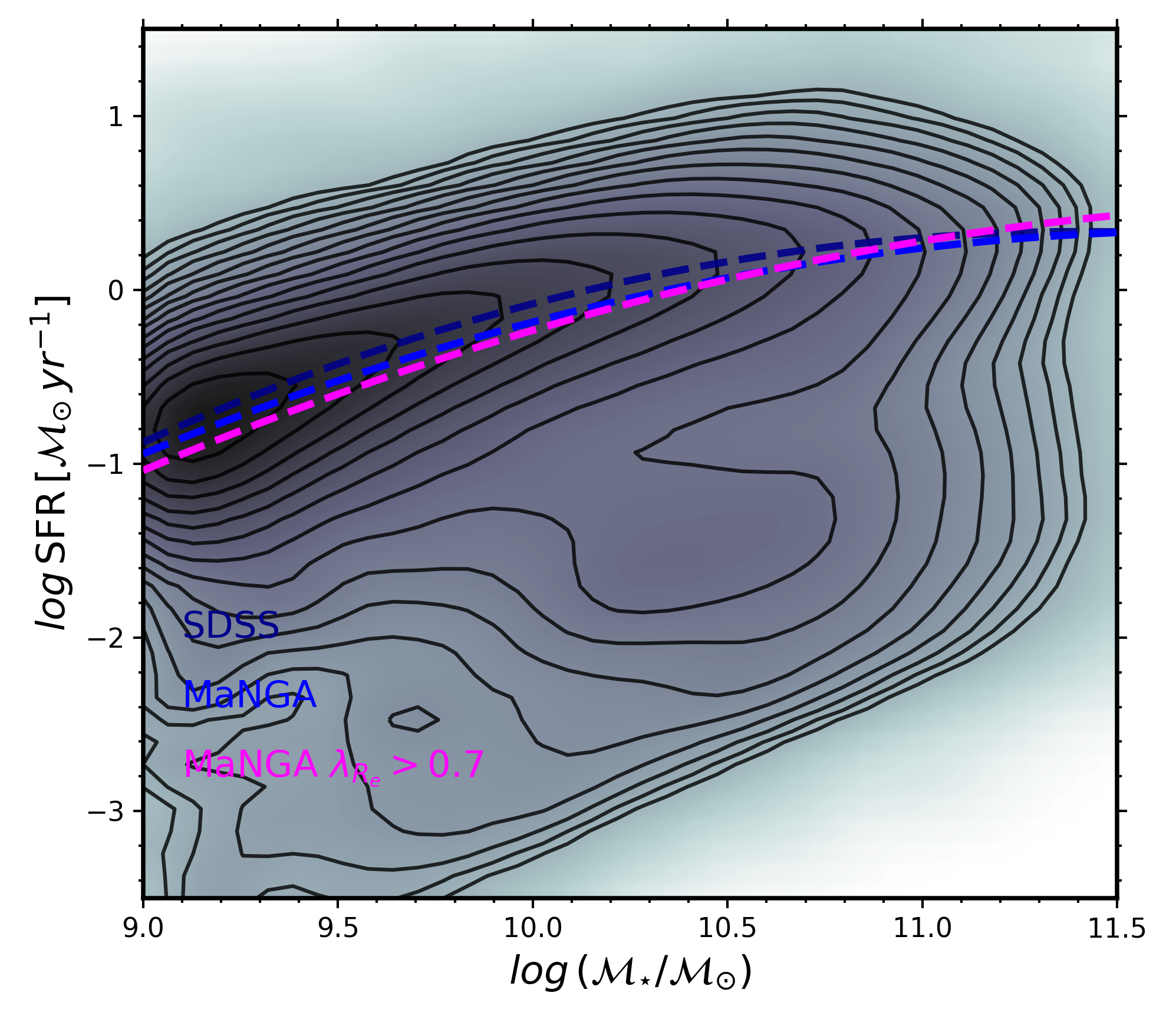}
    \caption{
    The $V_{\mathrm{max}}$ corrected probability density distribution and its ridge (the dark blue dashed line labeled as "SDSS") of SDSS galaxies in sample $\mathcal{S}_{\mathrm{GSWLC}}$ in redshift range $0.01<z<0.08$ on $\mathrm{SFR}-\mathcal{M}_{\star}$ plane.
    Gray scale reflects the contour level with a decreasing order from dark to light.
    The SFMS defined by $\mathcal{S}_{\mathrm{MaNGA}}$ (lebeled by "MaNGA") and its subsample of fast rotating galaxies with $\lambda_{R_e}>0.7$ (lebeled by "MaNGA $\lambda_{R_e}>0.7$") is shown for comparison by blue and magenta dashed line separately.
        }
    \label{fig:sfms_appendix}
\end{figure}

Following \citet{2015ApJ...801L..29R}, the star formation main sequence (the dark blue dashed line in \autoref{fig:sfms_appendix} labeled as "SDSS") used in this work is defined by the ridge line of probability density distribution (illustrated by gray scale and contours) of SDSS galaxies in sample $\mathcal{S}_{\mathrm{GSWLC}}$ in redshift range $0.01<z<0.08$.

Probability density is measured via kernel density estimation with circular gaussian kernel used and its width determined by Scott's rule \citep{Scott2015Multivariate}.
Axes of $\mathrm{SFR}$ and $\mathcal{M}_{\star}$ are binned at resolution about 0.12 and 0.06 dex respectively.
And in the estimation each galaxy is weighted by $V_\mathrm{total}/V_\mathrm{max}$, the co-moving volume ratio between total volume in our redshift range and the volume in range $0.01<z<z_\mathrm{max}$ where $z_\mathrm{max}$ is the maximum redshift at which a certain galaxy can still be included in SDSS spectroscopic survey (given the r band apparent magnitude limit 17.77), or $z_\mathrm{max}=0.08$ when the maximum redshift is larger than 0.08.

Then the ridge line is located by searching for the maxima in each $\mathcal{M}_{\star}$ bin.
We fit it with a second-order polynomial and get:
\begin{equation}\label{equa:sfms}
\mathrm{log}\,\mathrm{SFR}=-26.8+4.76\,\mathrm{log}\,\mathcal{M}_{\star}-0.21\,(\mathrm{log}\,\mathcal{M}_{\star})^2
\end{equation}
And this analytic line is used throughout the work.

Just for comparison, the SFMS defined in the same way by using much smaller sample $\mathcal{S}_{\mathrm{MaNGA}}$ and its subsample of fast rotating galaxies with $\lambda_{R_e}>0.7$ is shown by blue and magenta dashed line separately.
The difference is markedly small.

\section{Reproduction of Fig. 1 with SFR measured in effective radius}\label{app:repro}

\begin{figure*}
    \includegraphics[width=0.45\textwidth]{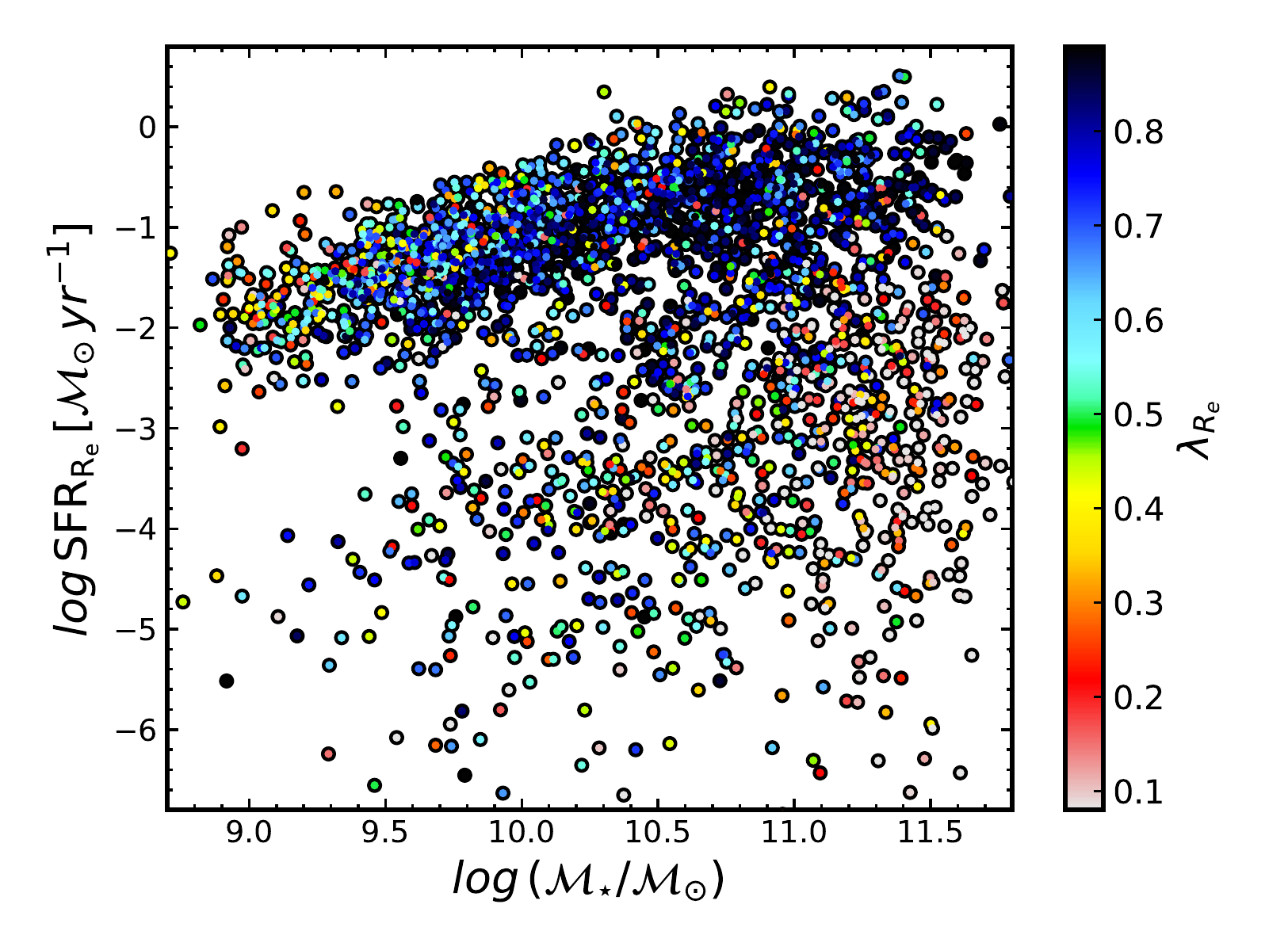}
    \includegraphics[width=0.45\textwidth]{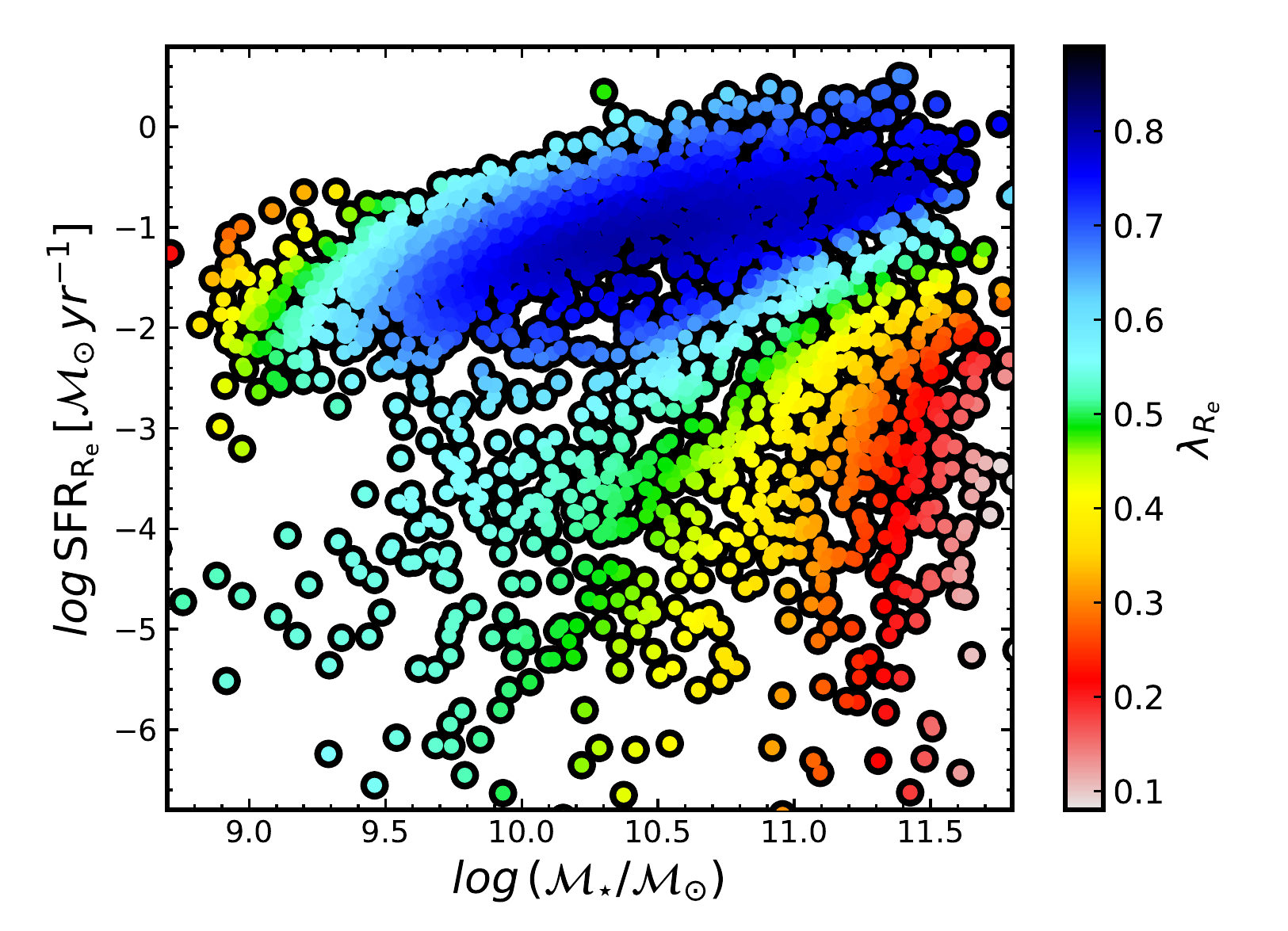}
    \caption{Reproduction of \autoref{fig:sfrm} using SFR measured from the MaNGA IFS data, using the H$\alpha$ fluxes within one half-light radius.
        }
    \label{fig:repro}
\end{figure*}

In this section we test the robustness of the qualitative trend we presented in \autoref{fig:sfrm}, by using a completely different approach to infer the SFR. For this we use the SFR derived using the H$\alpha$ fluxes within one effective radius $\mathrm{SFR}_{R_e}$ in the MaNGA data. Our alternative version of \autoref{fig:sfrm}, when using the MaNGA SFR is shown in \autoref{fig:repro}.

To obtain ${\rm SFR}_{R_e}$, we took MAPS files (the ones used for measuring the spin parameter) for MaNGA galaxies in our sample, as produced by the MaNGA DAP \citep{Westfall2019}. The approach used for the extraction of the gas fluxes from the MaNGA cubes was described in \citet{Belfiore2019}.
For spaxels that fall into one elliptical effective radius (defined by r-band elliptical Petrosian effective radius, axial ratio and position angle catalogued in the NSA), we used BPT diagram \citep{1981PASP...93....5B} to classify them into star-forming and non-star-forming according to the definition of \citealt{2003MNRAS.346.1055K}.
And then we corrected the $H\alpha$ flux of star-forming spaxels for dust extinction assuming a \citet{1983MNRAS.203..301H} Galactic extinction curve with $R_V=3.1$ and intrinsic $H\alpha/H\beta=2.86$.
Lastly we summed all dust-corrected $H\alpha$ flux of star-forming spaxels within one $R_e$ and converted it to SFR using the calibration in \citealt{2013seg..book..419C}:
\begin{equation}\label{hasfr}
\mathrm{SFR}_{H\alpha}/(\mathcal{M}_{\odot}\,yr^{-1})=5.5\times10^{-42}\times L_{{H\alpha}}/(\mathrm{erg}\,\,s^{-1})
\end{equation}
A factor of 95\% was then applied to adjust from Kroupa to Chabrier IMF \citep{2007ApJS..173..267S}.

\section{Testing how the distributions on $\lambda-\epsilon$ plane change with inclination}\label{app:inc}

In this appendix, we show how the distribution on $\lambda-\epsilon$ plane of a sample of models change with their inclination angles.
Such that we prove star-bursting MaNGA galaxies are not merely the face-on version of star-forming MaNGA galaxies.

\begin{figure}
  \includegraphics[width=0.47\textwidth,trim={0 0cm 0 0},clip]{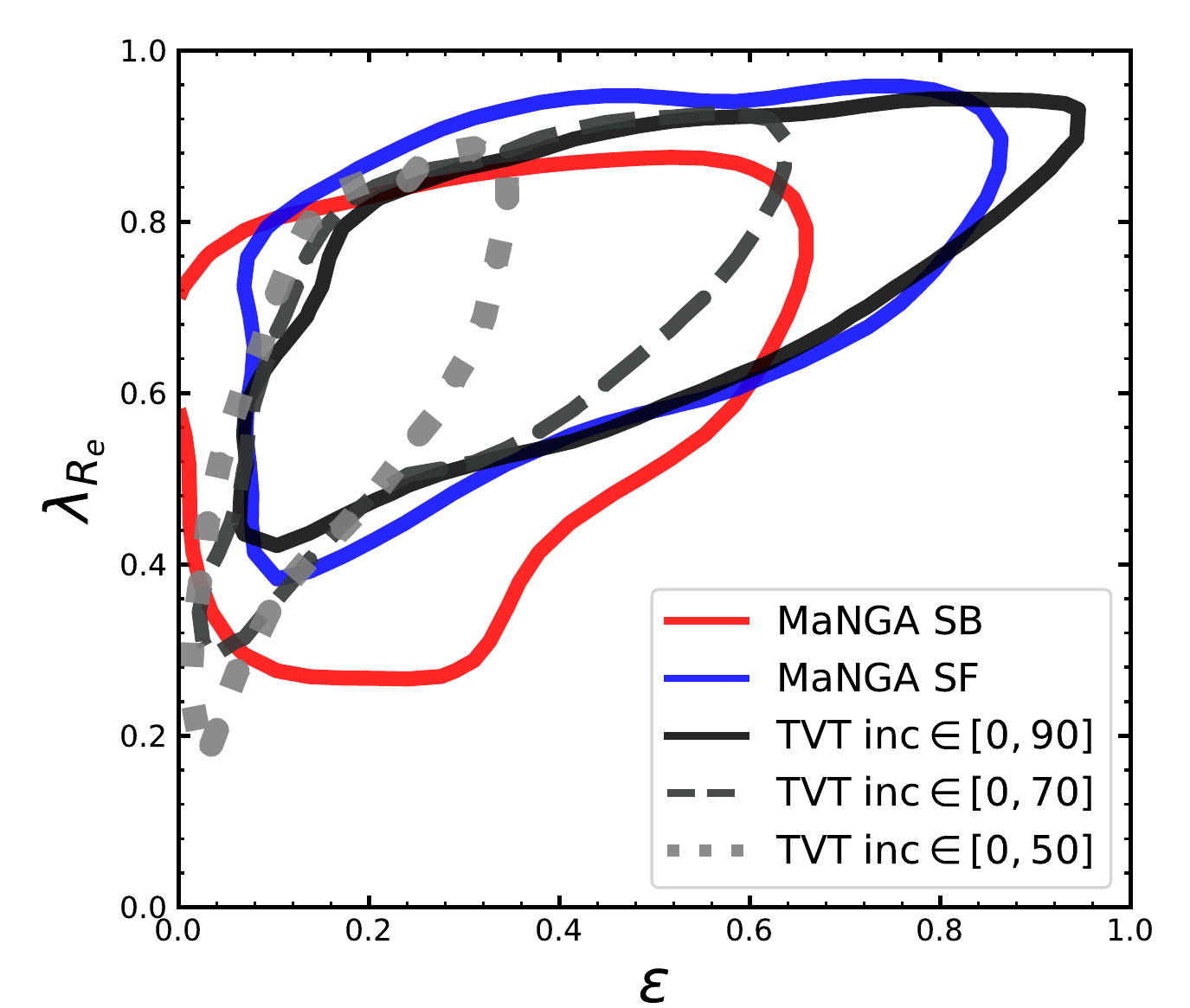}
    \caption{Probability density contours enclosing 68\% of total for MaNGA star-forming (blue), MaNGA star-bursting (red) galaxies and models based on tensor virial theorem (TVT) with different inclination restrictions.
    }
    \label{fig:inc}
\end{figure}

In \autoref{fig:inc}, the red and blue line denote the probability density contour enclosing 68\% of total, respectively for the star-bursting and star-forming galaxies in our MaNGA sample in stellar mass range $10^{9}-10^{11} \mathcal{M}_{\odot}$.
The probability density distributions are derived in the same manner as in the main text.
And the stellar mass bin is chosen to be the range for which we observe the drop of $\lambda_{R_e}$ when going beyond the SFMS.
Compared with star-forming galaxies it is clear that the distribution of star-bursting galaxies populates lower left, consistent with their lower $\lambda_{R_e}$ on average.
However, it may not be trivial to convince one that this is due to their smaller intrinsic ellipticity rather than lower inclination angles.
As when cutting off the high inclination tip of the star-forming distribution, the remaining distribution may resemble the star-bursting one.

To investigate this we created a sample of 3000 model galaxies based on tensor virial theorem.
For each model the intrinsic ellipticity $\epsilon_\mathrm{intr}$ was randomly drawn from a Gaussian distribution with mean 0.95 and standard deviation 0.25 trimmed between $(0,1)$, and anisotropy uniformly taken from $[0,0.7\times \epsilon_\mathrm{intr}]$ and with a random orientation.
So that, as shown by the black solid line in \autoref{fig:inc}, we get a density contour of model galaxies similar to the MaNGA star-forming one.
We further assembled two samples of models with the same parameter configurations but restricted in certain inclination range $[0,70]$ and $[0,50]$ degrees.
Their density contours enclosing 68\% of total are illustrated by black dashed and gray dotted line separately.

The result shows that when restricting models to be more face-on, the distributions still reach the high $\lambda_{R_e}$ envelope of the distribution without restriction.
This will be a key discrepancy if one thinks the star-bursting galaxies are only more face-on than the star-forming galaxies.
Because the star-bursting galaxies have lower $\lambda_{R_e}$ not only due to their distribution having lower bottom boundary but also the lower top boundary, suggesting intrinsically lower ellipticity.

\section{Images and kinematic maps of the "stragglers"}\label{app:straggler}

In this appendix, from the perspective of line-of-sight velocity and velocity dispersion maps, we show that those "stragglers" on $\lambda_{R_e} - \epsilon$ plane, i.e. galaxies with star formation level mismatched with their measured $\lambda_{R_e}$ discussed in \autoref{subsec:general}, are mostly galaxies with significant counterrotating stellar components.

\begin{figure*}
    \includegraphics[width=0.84\textwidth,trim={3cm 6.5cm 3cm 5cm},clip]{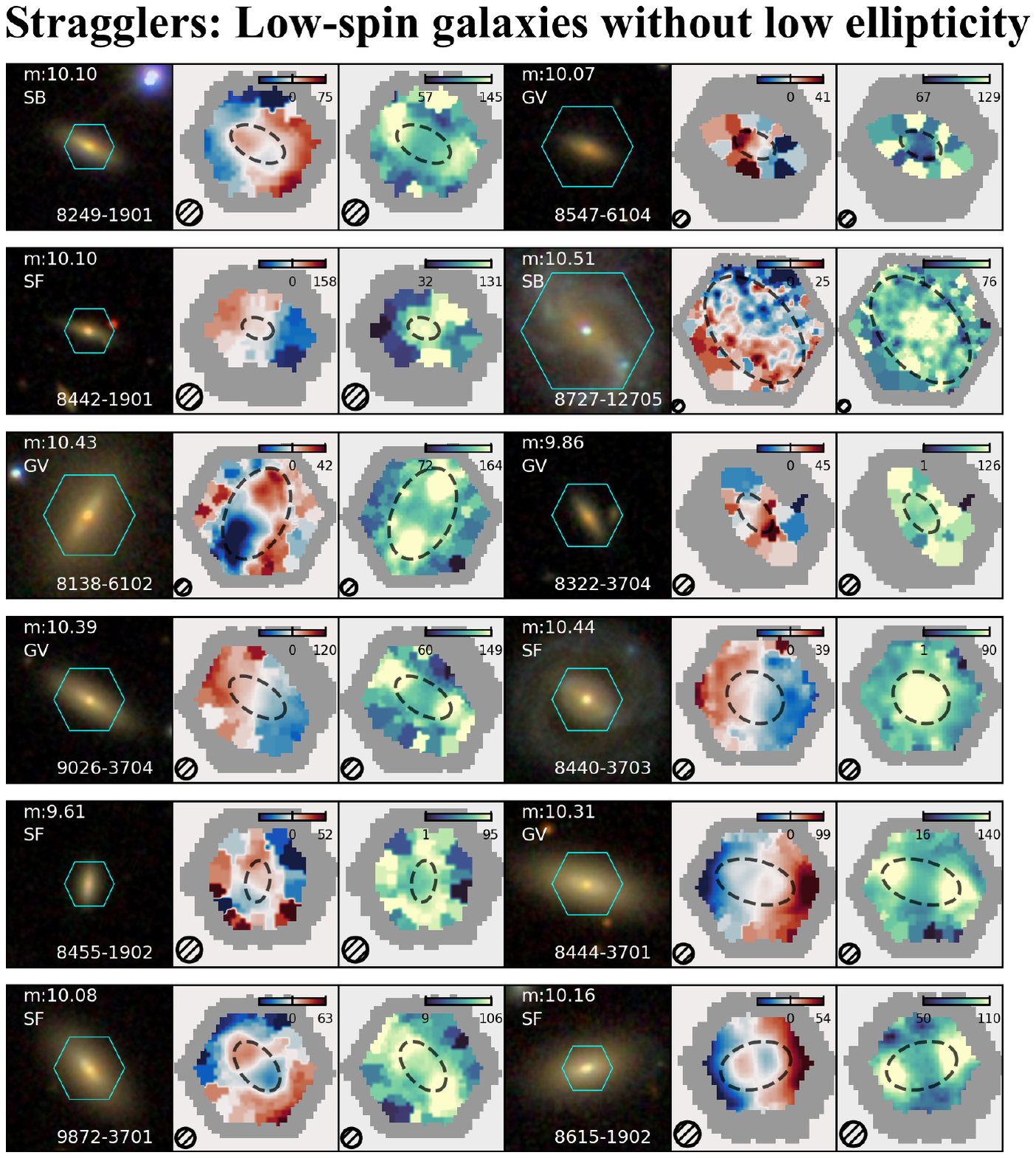}
    \caption{Images and kinematic maps of the 12 "stragglers".
    For each galaxy, the panels from left to right represent, respectively (1) SDSS g-r-i composite image with MaNGA bundle superimposed, (2) line-of-sight velocity and (3) line-of-sight velocity dispersion.
    In each image, the MaNGA plate-ifu ID for the galaxy is shown at the bottom together with its stellar mass and star formation level (SB, SF, GV corresponding to $\Delta_{\mathrm{MS}}>0.35$, $-0.35<\Delta_{\mathrm{MS}}<0.35$, $-1.35<\Delta_{\mathrm{MS}}<-0.35$) shown on the top left.
    For velocity and dispersion maps, FWHM of MaNGA beam is denoted at bottom left.
    Colour bars indicate the dynamic range in unit km/s and note that the range of velocity map is symmetric with respect to zero.
    Black dashed ellipses mark the boundary within which the spin parameter $\lambda_{R_e}$ is measured.
    }
    \label{fig:maps}
\end{figure*}

Like already mentioned in the main text, the maps shown in \autoref{fig:maps} are for 12 galaxies in the two lower mass bins in the box $0<\lambda_{R_e}<0.3\ \& \ 0.4<\epsilon<0.6$ with $\Delta _{\mathrm{MS}}>-1$.
For each galaxy, from left to right its SDSS g-r-i composite image, line-of-sight velocity map and line-of-sight dispersion map is displayed respectively.
Noteworthily, many of them do not have clear hourglass-like velocity field and centrally peaked dispersion field, both of which are features of normal rotators.
Instead, there are a large fraction of velocity maps indicating flipped rotating axis (e.g. the first, fifth and sixth of the second column) and also dispersion maps with two peaks away from center (the so-called "two-sigma" feature).
These are signals that galaxies have significant counterrotating stellar components.
By counting the "two-sigma" feature in dispersion maps, which is most sensitive to counterrotating disks, here we identify at least 7 out of 12 are counterrotating galaxies.

\section{Low-spin galaxies on the SFMS}\label{app:lowspinsf}

Galaxies on the SFMS are generally fast rotating while there are still systems with apparently low values of $\lambda_{R_e}$.

\begin{figure*}
    \includegraphics[width=0.93\textwidth,trim={1.2cm 5.cm 1.2cm 0.2cm},clip]{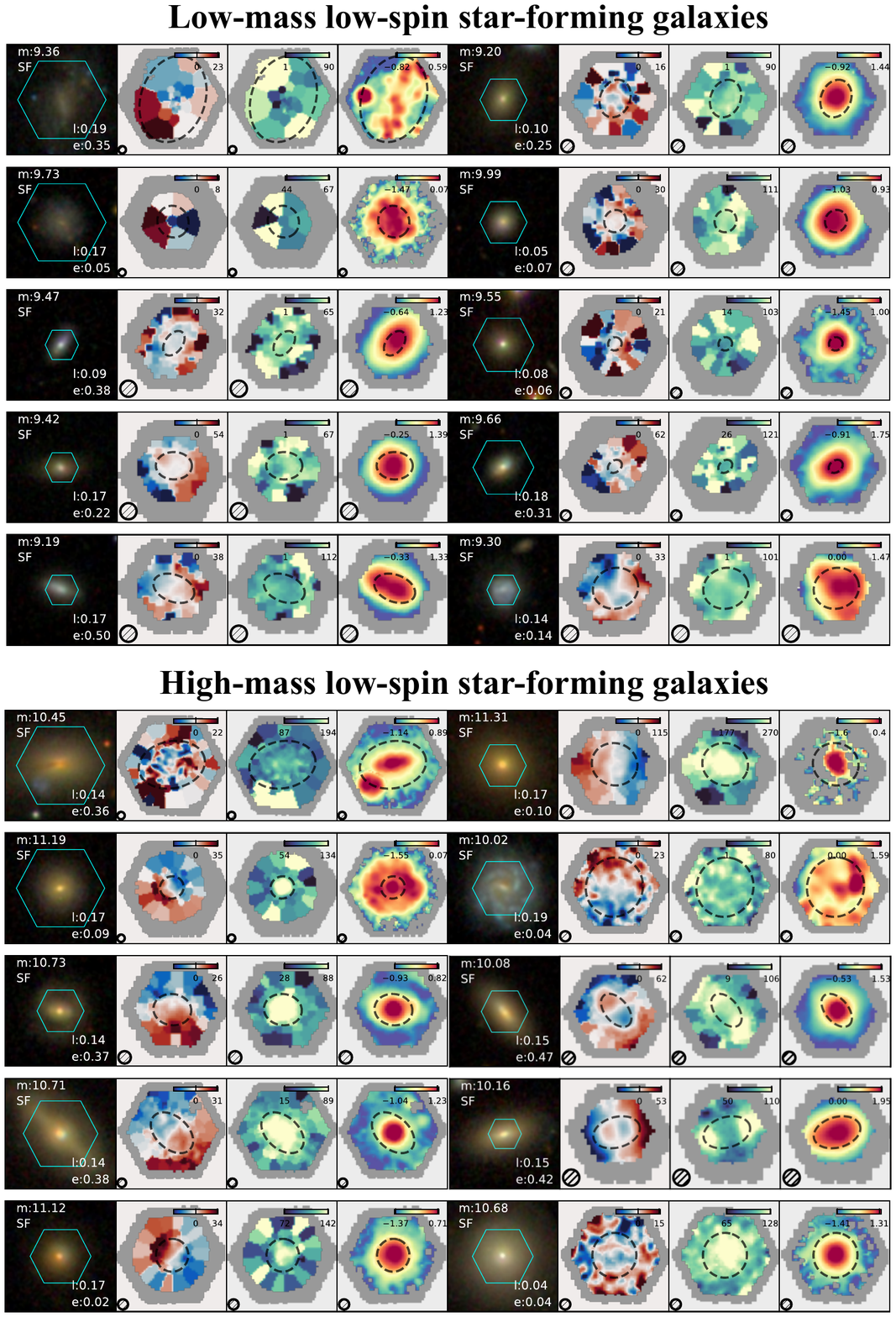}
    \caption{Low-spin ($\lambda_{R_e}<0.2$) star-forming galaxies with low mass (upper section; $10^{9}-10^{10} \mathcal{M}_{\odot}$) and high mass (lower section; $10^{10}-10^{11.5} \mathcal{M}_{\odot}$).
    Additional to the maps shown in Fig. \ref{fig:maps}, here we also show $H\alpha$ flux density maps in $\mathrm{erg}/(s \, \mathrm{cm}^2 \, \mathrm{spaxel})$ in log scale at the end of row for each galaxy.
    The value of $\lambda_{R_e}$ and ellipticity are attached at the bottom of SDSS images.
    }
    \label{fig:maps1}
\end{figure*}

We have visually checked the velocity and dispersion maps of all galaxies on the SFMS with $\lambda_{R_e}<0.2$, in low ($10^{9}-10^{10} \mathcal{M}_{\odot}$; 24 galaxies) and high stellar mass ($10^{10}-10^{11.5} \mathcal{M}_{\odot}$; 16 galaxies) bins respectively.
A randomly chosen sample of 10 galaxies of each sort are shown in \autoref{fig:maps1}.

As can be seen from the upper section, most of these low-mass galaxies display complex velocity field, suggesting they being intrinsically slowly rotating systems.
Many of them have low angular size but are still well resolved by the MaNGA IFU.
In the lower section, it shows that among these massive galaxies there are also two with complex velocity field (the first and the last).
But clearly, different from the low-mass systems, six of these massive galaxies display normal rotation velocity field (even though some of these galaxies are also small) and feature very open velocity contours which is the signal of face-on disks (see also Fig. 3 of \citet{2016ARA&A..54..597C} for an illustration).
A representative is the first galaxy in the right column.
This suggests a significant fraction of them have apparently low spin primarily because they are viewed relatively face-on.
We note that the apparent red colour of these massive star-forming galaxies is partly due to their lower specific SFR and higher dust attenuation than star-forming galaxies of lower mass.
And five of these ten massive galaxies, for which we collected GALEX NUV and WISE 24 microns data, are all on the SFMS according to SFRs derived from NUV plus 24 microns luminosity.

This difference between low- and high-mass galaxies accords the decreasing $\lambda_{R_e}$ trend toward low mass end that we see on the SFMS in \autoref{fig:sfrm}.

\bsp    
\label{lastpage}
\end{document}